\documentclass[sigconf, nonacm]{acmart}

\newcommand\vldbdoi{XX.XX/XXX.XX}
\newcommand\vldbpages{XXX-XXX}
\newcommand\vldbvolume{14}
\newcommand\vldbissue{1}
\newcommand\vldbyear{2020}
\newcommand\vldbauthors{\authors}
\newcommand\vldbtitle{\shorttitle} 
\newcommand\vldbavailabilityurl{URL_TO_YOUR_ARTIFACTS}
\newcommand\vldbpagestyle{plain}

\usepackage[prologue,dvipsnames]{xcolor} 
\usepackage{xpunctuate}
\usepackage{caption}
\usepackage{makecell}
\usepackage{subcaption}
\usepackage{amsmath}
\usepackage{pifont}
\usepackage{cleveref}
\usepackage{enumitem}
\usepackage{algorithm}
\usepackage{algorithmic}

\usepackage{graphicx}
\usepackage{array}
\usepackage{booktabs}
\usepackage{multirow}
\usepackage{colortbl}
\usepackage{geometry}
\usepackage{hyperref}

\usepackage{ulem}

\newcommand{\rot}[1]{\rotatebox{90}{\parbox{2cm}{\centering #1}}}

\newcommand{\ie}{{i.e.,}\xspace}
\newcommand{\eg}{{e.g.,}\xspace}
\newcommand{\etal}{{et~al\xperiod}\xspace}

\newcommand{\etc}{{etc\xperiod}\xspace}
\newcommand{\vs}{{vs\xperiod}\xspace}
\newcommand{\wrt}{{w.r.t.}\xspace}

\newcommand{\cmark}{\textcolor{ForestGreen}{\ding{51}}}%

\begin{document}
\title{Task-Centered Benchmark for Interactive Network Visualization \& Analysis}
\subtitle{Experiments \& Analysis}

\author{Ameya Patil}
\orcid{0000-0002-9743-4264}
\affiliation{%
  \institution{University of Washington}
  \city{Seattle}
  \country{USA}
}
\email{ameyap2@cs.washington.edu}

\author{Wei Jun Tan}
\orcid{0009-0004-1881-3453}
\affiliation{%
  \institution{University of Washington}
  \city{Seattle}
  \country{USA}
}
\email{wj428@cs.washington.edu}

\author{Ishan Sinha}
\orcid{0000-0002-4427-2708}
\affiliation{%
  \institution{University of Washington}
  \city{Seattle}
  \country{USA}
}
\email{ishans2@cs.washington.edu}

\author{Leilani Battle}
\orcid{0000-0003-3870-636X}
\affiliation{%
  \institution{University of Washington}
  \city{Seattle}
  \country{USA}
}
\email{leibatt@cs.washington.edu}

\begin{abstract}

    Interactive network visualization and analysis (INVA) enables iterative, visual and algorithmic analysis of large network datasets. Although numerous benchmarks have been developed to evaluate different graph analysis algorithms and systems, we observe a lack of such efforts for interactive network data understanding. In this work, we address the question - \textbf{\textit{How well do existing graph systems serve the purpose of Interactive Network Visualization and Analysis?}} To this end, we build and demonstrate the use of the first task-centered benchmarking framework to evaluate a variety of graph system backends on INVA workloads. Our benchmarking results highlight a gap between both the capabilities and performance of existing graph systems for INVA use cases, and uncover possible bugs in these systems. Based on our benchmarking results, we reveal new opportunities for research and development to better support interactive network visualization and analysis.

\end{abstract}

\maketitle

\pagestyle{\vldbpagestyle}
\begingroup\small\noindent\raggedright\textbf{PVLDB Reference Format:}\\
\vldbauthors. \vldbtitle. PVLDB, \vldbvolume(\vldbissue): \vldbpages, \vldbyear.\\
\href{https://doi.org/\vldbdoi}{doi:\vldbdoi}
\endgroup
\begingroup
\renewcommand\thefootnote{}\footnote{\noindent
This work is licensed under the Creative Commons BY-NC-ND 4.0 International License. Visit \url{https://creativecommons.org/licenses/by-nc-nd/4.0/} to view a copy of this license. For any use beyond those covered by this license, obtain permission by emailing \href{mailto:info@vldb.org}{info@vldb.org}. Copyright is held by the owner/author(s). Publication rights licensed to the VLDB Endowment. \\
\raggedright Proceedings of the VLDB Endowment, Vol. \vldbvolume, No. \vldbissue\ %
ISSN 2150-8097. \\
\href{https://doi.org/\vldbdoi}{doi:\vldbdoi} \\
}\addtocounter{footnote}{-1}\endgroup

\ifdefempty{\vldbavailabilityurl}{}{
\vspace{.3cm}
\begingroup\small\noindent\raggedright\textbf{PVLDB Artifact Availability:}\\
The source code, data, and/or other artifacts have been made available at \href{https://github.com/WeiJun428/graph-system-benchmark}{https://github.com/WeiJun428/graph-system-benchmark} and\\ \href{
https://osf.io/3z4hk/overview?view_only=21e3de8aea404b1186c4d02938c24996}{https://osf.io/3z4hk}
\endgroup
}

\section{Introduction}

The ever-growing importance of large-scale network data has sparked increased interest in visualizing and analyzing large networks~\cite{bonifati_roadmap_2024}. In response, many graph algorithms~\cite{tripathy_scalable_2018, wu_graph_2023, rodriguez_clustering_2019, meng_survey_2025}, graph processing frameworks and systems~\cite{malewicz_pregel_2010, low_graphlab_2012, roy_x-stream_2013, xin_graphx_2013} have been developed to support large graph analytics. However, these solutions are predominantly opaque to domain experts or non-programmers, and fail to account for \emph{human-in-the-loop or interactive understanding of network data}.

In this work, \textbf{we focus on a relatively understudied aspect of network data understanding: \emph{interactive} network visualization and analysis (INVA).} INVA facilitates quick exploration and sensemaking of large complex network datasets including critical tasks across the network analysis pipeline such as data import, visualization and algorithmic processing, and exporting visualizations for information dissemination, commonly performed through intuitive graphical user interfaces~\cite{sinva_survey_2026}. As has been shown in prior work in benchmarking for relational data~\cite{battle_crossfilter_2020, eichmann_idebench_2020, battle_database_2020}, interactive visualization and analysis workloads can trigger distinct queries and bursty frequencies, which existing graph analytics benchmarks do not simulate~\cite{iosup_ldbc_2020, erling_ldbc_2015}. Individual graph algorithms used within such workloads have been evaluated for efficacy, speed and resource utilization only in a standalone manner~\cite{meng_revisiting_2025, rodriguez_clustering_2019, wu_graph_2023, ribeiro_survey_2022, dhulipala_graph_2020, nobre_state_2019, filipov_are_2023}, and not in the context of INVA.
Thus there is a need to evaluate graph systems comparatively \textit{in the context of human-in-the-loop or interactive network visualization and analysis scenarios}.

In this work, we answer the overarching question - \textbf{\textit{how well do existing graph systems support interactive network visualization and analysis (INVA)?}} Towards this end, we (1) develop a task-centered model of how analysts perform INVA to generate representative INVA workloads, and (2) design and implement a task-centered benchmarking framework to measure how well existing graph systems perform on these INVA workloads over large-scale networks. We demonstrate the use of our benchmarking framework by evaluating the backends of a set of prominent INVA and graph database systems. Through the results of the benchmarking evaluation, \textbf{we reveal a gap both in the kind of analyses supported by these systems, and their performance for INVA use cases, while also exposing correctness issues in some of our evaluated graph systems.} We conclude with a discussion of new opportunities for research and development in Interactive Network Visualization and Analysis.

Our benchmark redefines graph system performance in the context of human-in-the-loop scenarios, shifting away from abstract complexity measures and towards measuring how these systems advance (or hinder) human insight. To the best of our knowledge, \textbf{our work is the first to build a comprehensive benchmark for human-in-the-loop network data understanding.} We view this work as an initial performance study over network data understanding systems for large networks, demonstrating the value of developing benchmarks for INVA, which can help researchers develop and evaluate new graph visualization and analysis systems. In summary, we make the following contributions:

\begin{enumerate}

    \item We develop a \textbf{task-centered model of Interactive Network Visualization and Analysis (INVA)} to generate representative INVA workloads.

    \item We design and implement a \textbf{task-centered benchmarking framework} to evaluate network data understanding systems for INVA workloads on large-scale network data.
    
    \item We present \textbf{results from benchmarking prominent INVA and graph database systems} supporting INVA use cases, which reveal new avenues for future database research.
    
\end{enumerate}

For the rest of the paper, we adopt the same terminology as~\cite{sinva_survey_2026} where `\textit{network}'
refers to real-life data which carries with it the context or semantics of the data, and `\textit{graph}' refers to the mathematical model or abstraction that is used to enable analysis of network data. A \textit{network} has \textit{nodes} and \textit{links}, while a \textit{graph} has \textit{vertices} and \textit{edges}.

\section{Benchmarking Framework}
\label{sec:benchmarking-framework}

Prior benchmarking work for graph systems has primarily focused on graph databases and graph processing systems for individual graph algorithms run in a \textit{standalone} manner~(\cref{sec:bg-benchmarking}). Our aim is to address the lesser explored needs of human analysts of network data, who perform a \textit{sequence of graph operations} to understand the network data. We thus present our task-centered benchmarking framework for interactive network visualization and analysis (INVA) of large-scale networks. To the best of our knowledge, this is the first benchmark addressing systems for human-in-the-loop understanding of network data.

\begin{figure}[ht]
    \centering
    \includegraphics[width=\linewidth, trim={7em 9em 5em 9em}, clip]{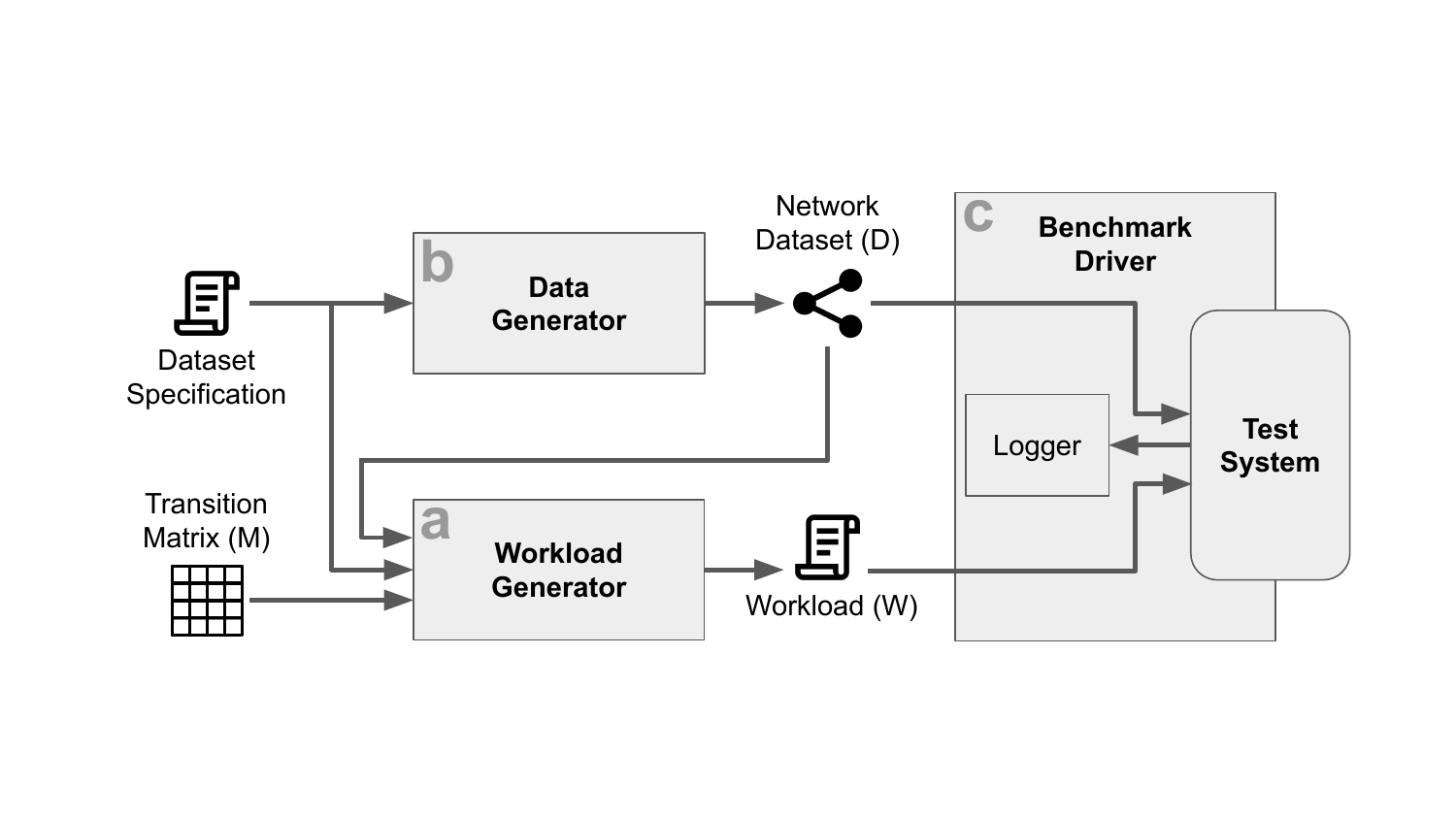}
    \caption{Architecture of the benchmarking framework}
    \label{fig:benchmark-architecture}
    \Description[Block diagram of the benchmarking framework]{Block diagram of the benchmarking framework}
\end{figure}

Our benchmarking framework is designed based on prior work in graph task taxonomies, graph benchmarking and benchmarking for interactive data analytics, as described in~\cref{sec:related-work}. It consists of 3 modules: (a) the workload generator, (b) the data generator, and (c) the benchmark driver, of which the workload generator module is the main contribution. \Cref{fig:benchmark-architecture} shows how these individual modules work together in the framework. We now describe some terminologies used in our framework followed by the design and implementation of each module.

\subsection{Base Terminologies}

We define the following basic terms in the context of INVA to build upon further and describe the benchmarking framework:

\subsubsection{Dataset $(D)$} Network data having a set of nodes $N$ and a set of links $L$, with attribute sets $Na$ and $La$ respectively. A dataset may also be expressed using the relational model as $N(na_1, na_2, ..., na_p)$ and $L(la_1, la_2, ..., la_q)$, where $na_i \in Na$ and $la_i \in La$ are the node and link attributes respectively.

\subsubsection{Interaction $(i)$} An analysis or visualization operation performed on the dataset $D$, optionally taking in arguments, and generating a result, a set of results, or a visualization as output.
    
\subsubsection{Workflow $(w)$} The dataset $D$ or its subset, and a sequence of interactions $I = <i_1, i_2, ... i_n>$ performed on the data, representing an analysis approach or goal (blue box in~\cref{fig:inva-model}).
    
    
\subsubsection{Workload $(W)$} A sequence of workflows $<w_1, w_2, ... w_m>$ where each \textit{workflow} represents a potentially different analysis approach or goal (gray box in~\cref{fig:inva-model}). The entire workload thus represents different approaches of understanding the data.

\subsection{Workload Generator Module}
\label{sec:workload-generator-module}


An important aspect of benchmarking is the creation of workloads on which to test the systems. Prior work in benchmarking for interactive analytics~
\cite{eichmann_idebench_2020, battle_database_2020} performed user studies to create a model of interactive analysis of relational data and generate corresponding workloads. However, such models are better informed by the analyses performed by domain experts rather than by crowdsourced users~\cite{wu_rational_2023, huang_vistruct_2025}. 
Thus, instead of resorting to crowdsourced studies, we build our model of INVA workloads~(\cref{sec:inva-model}) based on prior research in graph task taxonomies~\cite{nobre_state_2019, sinva_survey_2026, gathani_grammarbased_2022}, and graph analysis use cases by domain experts~\cite{chen_understanding_2020, ocallaghan_uncovering_2013, brandon_graph_2020, bodin_role_2009, wojcik_social_2021, charitou_biological_2016, wang_data-driven_2021, azizi_visibility_2025, kuikka_network_2025, hadaj_plans_2022, borsboom_network_2021, kalhor_insight_2022}. We generate our workloads using this INVA model~(\cref{sec:inva-workload-generator}).


We first present our INVA model which captures how humans visualize and analyze network data to understand it, and then describe our INVA workload generator (\cref{fig:benchmark-architecture}a) which uses this model.

\subsubsection{INVA Model}
\label{sec:inva-model}

\begin{figure}[ht]
    \centering
    \includegraphics[width=\linewidth,trim={2.5em 5em 2.5em 6em},clip]{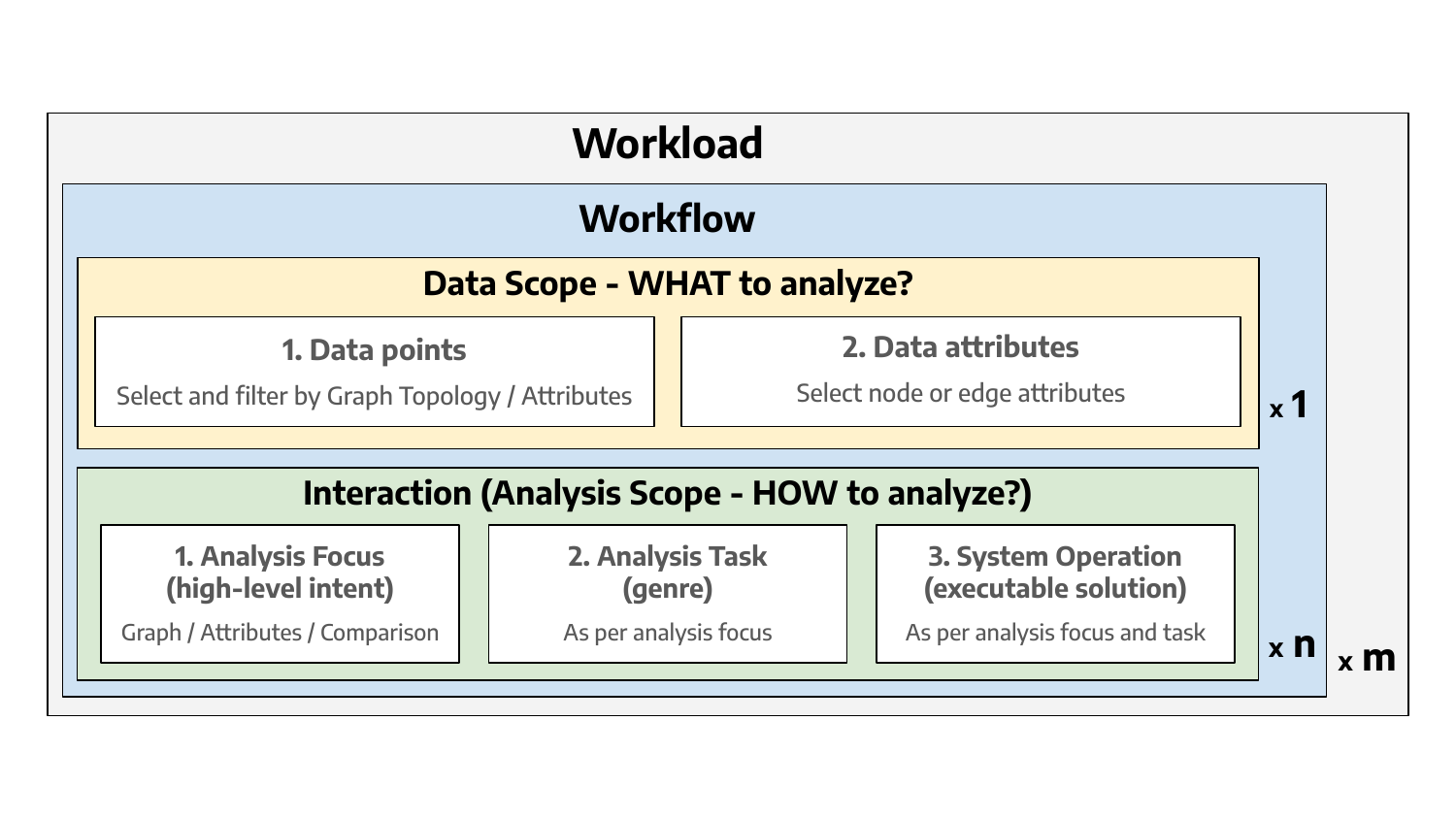}
    \caption{Characterizing INVA workflow and workload. Each workflow has a \textit{data scope} and \textit{n} interactions. \textit{m} such workflows are concatenated to form a workload.}
    \label{fig:inva-model}
    \Description[A nested diagram showing how INVA workloads are composed of workflows, data scope and analysis scopes]{A nested diagram showing how INVA workloads are composed of workflows, data scope and analysis scopes}
\end{figure}

To generate interactive network visualization and analysis workloads for benchmarking purposes, we first need a formal model of interactive network visualization and analysis, to characterize how humans analyze network data. We begin by defining two terms in this regard.


\begin{enumerate}[parsep=2pt]

\item \textbf{Data Scope:} The subset $D_j$ of the network data, being analyzed during a workflow $w_j$, such that all the interactions within the workflow are focused on this subset (\cref{fig:inva-model} yellow box). It is characterized by (1) the \textit{data points} $(dp)$ \ie a subset of the nodes/links, or rows of the node/link table, and (2) the \textit{data attributes} $(da)$ \ie a subset of the node/link attributes, or columns of the node/link table. Thus,
$$ D_j \subseteq D \text{ for workflow } w_j$$ 
$$dp(D_j) = (N_j, L_j)\text{, where }N_j \subseteq N ~\&~ L_j \subseteq L$$
$$da(D_j) = (Na_j, La_j)\text{ where }Na_j \subseteq N_a ~\&~ La_j \subseteq L_a$$

\item \textbf{Analysis Scope:} The characterization of an interaction performed within a workflow (\cref{fig:inva-model} green box). This characterization is done in a hierarchical manner moving from a high-level intent behind performing the analysis (analysis focus) to a low-level executable solution for the analysis (system operation).

\textbf{Analysis Focus ($af$):} A high-level intent of analyzing the data scope during each interaction in a workflow.

\textbf{Analysis Task ($at$):} A genre of analysis performed under each analysis focus, where each genre corresponds to a semantic grouping of analysis types. This semantic grouping helps us simulate a realistic and semantically meaningful INVA workload, as detailed in~\cref{sec:inva-workload-generator}.

\textbf{System Operation ($op$):} An executable solution which facilitates performing the analysis task. This represents the lowest level of abstraction in our INVA model, in terms of which we issue commands to the test system for benchmarking.

\end{enumerate}

Thus, an \textit{interaction} is expressed as
$$ i = \{af, at, op\}$$

We refer to prior work in graph task taxonomy (\cref{sec:bg-graph-task-taxonomies}) for the different kinds of analysis foci, analysis tasks and system operations. One of the earliest graph task taxonomies by Lee~\etal~\cite{lee_task_2006} categorizes graph tasks as (1) Topology based: adjacency, accessibility, common connection, connectivity; (2) Attribute based: filtering, distribution, range, compute derived attributes; (3) Browsing and (4) Overview. Nobre~\etal~\cite{nobre_state_2019} concisely taxonomize graph analysis tasks as analyzing either the topology, or the attributes of the network data. We use these two graph task taxonomies along with the survey of operational INVA tasks by~\cite{sinva_survey_2026} to create our hierarchical INVA model containing three different analysis foci, along with the constituent analysis tasks for each focus, and the constituent system operations within each analysis task. The hierarchical nature of our model is inspired from the work of Gathani~\etal~\cite{gathani_grammarbased_2022} following which, we map low level system operations to each high level analysis focus. Thus through our INVA model, we provide a tangible use case for graph task taxonomies and grammars, thereby proving their utility. We now describe these analysis foci and the constituent analysis tasks, along with some example system operations under each analysis task. \Cref{tab:inva-model-analysis-scope-characterization} lists only those system operations which are currently supported in our benchmarking framework.

\begin{enumerate}
    \item \textbf{Graph Focus:} Interactions with a graph focus are intended to understand the topology of the network, and require the use of information about \textbf{both} the nodes and links. We have the following analysis tasks under graph focus:
    
    \begin{enumerate}
        \item \textbf{Connections:} Tasks for moving between nodes~\eg shortest paths, network diffusion/flow, spanning trees,~\etc
        
        \item \textbf{Group:} Tasks for detecting or working with groups of nodes \eg clusters, connected components, k-core, pattern matching~\etc
        
        \item \textbf{Metrics:} Tasks for computing topology based node/link properties~\eg centrality measures, local clustering coefficient~\etc
        
        \item \textbf{Summary:} Tasks for computing statistics which summarize the topology~\eg triadic census, average path length, graph diameter~\etc
        
        \item \textbf{Visualize:} Tasks for visualizing topology~\eg graph layout (node-link) or node-ordering (adjacency matrix)
        
        \item \textbf{Edit:} Tasks for editing the network using both the nodes and links information~\eg add/delete node/link, add new node/link attribute using adjacent link/node information~\etc
    \end{enumerate}

    \item \textbf{Attribute Focus:} Interactions with an attribute focus are intended to understand the attributes of the network, and require the use of \textbf{either} the node attribute data \textbf{or} the link attribute data, but \textbf{not both together}. It consists of the following analysis tasks:
    
    \begin{enumerate}
        \item \textbf{Edit:} Tasks for editing either the node or link information in the network
        
        \item \textbf{Compute:} Tasks for computing attribute based node/link properties~\eg global aggregation of node/link attributes, transformation of node/link attributes, count~\etc
    \end{enumerate}

    \item \textbf{Comparison Focus:} Interactions with a comparison focus are intended to compare attributes for different nodes/links, or nodes/links with certain attribute values. This focus contains the following analysis tasks:
    
    \begin{enumerate}
        \item \textbf{Encode:} Tasks for encoding any node/link attribute in the graph visualization~\eg encode in color, size~\etc
        
        \item \textbf{Sort:} Tasks for sorting nodes/links based on attribute values
        
        \item \textbf{Correlate:} Tasks for gauging correlation between attributes \eg creating scatter plots, statistical tests~\etc
        
        \item \textbf{Extrema:} Tasks for gauging the distribution of node/link attributes~\eg  min/max value, outliers, histograms~\etc
    \end{enumerate}
    
\end{enumerate}

Apart from identifying different semantic genres of INVA, the analysis foci also separate different INVA genres based on data access patterns. \textit{Graph} focus accesses \textbf{both} node and link data, while \textit{attribute} and \textit{comparison} foci access \textbf{either} node \textbf{or} link data.

\begin{footnotesize}
    \begin{table}[t]
        \caption{Components of the INVA model (a) Data Scope, and (b) Analysis Scope. Each column represents a dimension along which we characterize the respective scope, while each row represents the possible options to characterize a scope along the concerned dimension.}
        \begin{subtable}[h]{\linewidth}
            \caption{Data Scope - What to analyze?}
            \label{tab:inva-model-data-scope-characterization}

            \begin{tabular}{>{\centering\arraybackslash}m{0.46\textwidth}>{\centering\arraybackslash}m{0.46\textwidth}}
            
                \toprule[1pt]
                
                \textbf{1. Data Points $(dp)$} (rows) & \textbf{2. Data Attributes $(da)$} (columns)\\
                \cmidrule[1pt](ll){1-1} \cmidrule[1pt](ll){2-2}

                \makecell[cc]{Select ($\sigma$)\\(k-hop neighborhood subgraph /\\Node attribute filter /\\Link attribute filter)} & \multirow{2}{*}{\makecell[bc]{\\ \\Project ($\Pi$)\\(Node attributes / Link attributes)}} \\

                \cmidrule[0.25pt](ll){1-1}

                \makecell[cc]{None\\(entire network is being analyzed)} & \\
    
                
                \bottomrule[1pt]
            \end{tabular}
        \end{subtable}

        \vspace{2mm}
        
        \begin{subtable}[h]{\linewidth}
            \caption{Analysis Scope - How to analyze?}
            \label{tab:inva-model-analysis-scope-characterization}

          \begin{tabular}{>{\centering\arraybackslash}m{0.15\textwidth}>{\centering\arraybackslash}m{0.15\textwidth}>{\centering\arraybackslash}m{0.6\textwidth}}
                \toprule[1pt]

                \textbf{1. Analysis Focus $(af)$} & \textbf{2. Analysis Task $(at)$} & \textbf{3. System Operations $(op)$} \\

                \cmidrule[1pt](ll){1-1} \cmidrule[1pt](ll){2-2} \cmidrule[1pt](ll){3-3}

                \multirow{5}{*}{\makecell[l]{Graph\\Focus}} & Connections & Path Existence Check / Shortest Path Finding / Single Source Shortest Path \\

                \cmidrule[0.25pt](ll){2-2} \cmidrule[0.25pt](ll){3-3}

                 & Group & Graph Clustering / Connected Components \\

                \cmidrule[0.25pt](ll){2-2} \cmidrule[0.25pt](ll){3-3}

                 & Metrics & Centrality Measures / Clustering Coefficient \\
                 
                 \cmidrule[0.25pt](ll){2-2} \cmidrule[0.25pt](ll){3-3}
                 
                 & Summary & Triadic Census / Graph Diameter / Avg Path Length / Global Clustering Coefficient \\

                \cmidrule[0.25pt](ll){2-2} \cmidrule[0.25pt](ll){3-3}

                 & Visualize & Graph Layout (node-link) / Node Ordering (adj matrix) \\

                \cmidrule[0.25pt](ll){2-2} \cmidrule[0.25pt](ll){3-3}

                 & Edit & Update nodes/links / Add node/link / Delete node/link \\

                \cmidrule[0.5pt](ll){1-3}

                \multirow{2}{*}{\makecell{Attributes\\Focus}} & Edit & Update node/link \\

                \cmidrule[0.25pt](ll){2-2} \cmidrule[0.25pt](ll){3-3}

                 & Compute & Custom Computation / Count \\

                \cmidrule[0.5pt](ll){1-3}

                 \multirow{4}{*}{\makecell{Comparison\\Focus}} & Encode & Encode attribute channel / Toggle Labels \\
                 
                \cmidrule[0.25pt](ll){2-2} \cmidrule[0.25pt](ll){3-3}
                 
                  & Sort & Sort node/link attribute \\

                \cmidrule[0.25pt](ll){2-2} \cmidrule[0.25pt](ll){3-3}
                  
                  & Correlate & Statistical Test / Scatterplot \\

                \cmidrule[0.25pt](ll){2-2} \cmidrule[0.25pt](ll){3-3}
                  
                  & Extrema & Minimum / Maximum / Distribution / Outliers \\
                
                \bottomrule[1pt]
          \end{tabular}      
        \end{subtable}

      \label{tab:inva-model}
    \end{table}
\end{footnotesize}

\subsubsection{INVA Workload Generator}
\label{sec:inva-workload-generator}

We now detail the implementation of the workload generator built using our INVA model. \Cref{alg:workload-generation} shows the pseudocode for the workload generation.

Each workflow $w_j$ is synthesized by choosing (1) a data scope $D_j$, and (2) a sequence of interactions $I_j$ pertaining to the data scope, in a top-down manner as per the hierarchy of our INVA model.

\vspace{1mm}
\noindent \textbf{(1) Data Scope Selection:} (lines 3-10) The \textit{data scope} $D_j$ is chosen in two steps:

    \begin{itemize}
    
        \item \textit{Data points} \ie the nodes/links to be analyzed, are chosen by performing a Select $(\sigma)$ operation (line 5) on the dataset. This includes using attribute filters for nodes or links, or fetching the k-hop neighborhood of certain nodes of choice. The result includes the induced subgraph of the filtered-in nodes/links. When no Select operation is performed to choose the data points ($\sigma$ = none), the entire network $D$ is being analyzed in the workflow. 
    
        \item \textit{Data attributes} \ie the node/link attributes to be analyzed in a data scope (equivalent to the Project $(\Pi)$ operation) are chosen from the set of both the node and link attributes (line 6). We sample at least 2 node/link attributes.
    
    \end{itemize}
    
Analysts may sometimes want to drill down into a subgraph of the previous data scope by applying more filters, or analyze a different set of attributes for the previous data scope. Thus, the data scope from the previous workflow may be reused to perform the Select and Project operations for the next workflow. We simulate this scenario by assigning probabilities to the choice of the original data $D$ (or previous data scope $D_{j-1}$), proportional to the node count of the filtered out data $|N| - |N_{j-1}|$ (or previous data scope $|N_{j-1}$|) (line 3). Analysts are less likely to drill down deeper into an already reduced dataset.

In both the choice of the Select operation and the data attributes, we use random selection. The use of more sophisticated selection techniques like choosing a data scope as per the exploration goals of the analyst~\cite{purich_adaptive_2025} is discussed in future work~\cref{sec:improving-benchmark}. \Cref{tab:inva-model-data-scope-characterization} summarizes the possible characterizations of the data scope currently supported in our benchmarking framework.

\vspace{1mm}
\noindent \textbf{(2) Interaction Sequence Generation:} (lines 13-26) We generate interaction sequences for a workflow such that each new interaction has a semantic dependency on both the previous interaction in the workflow, and the data scope of the workflow. Thus,
$$ i_k = f(i_{k-1}, D_j)\text{, for workflow }w_j$$


We model the dependency on the previous interaction using a Markov model with a transition probability matrix ($M$) to choose the next interaction given the previous interaction. Using this matrix, each interaction in the interaction sequence is generated as per the hierarchy of our proposed INVA model. First, the analysis focus is generated, followed by the analysis task and finally the system operation (lines 16-23). Workloads are created by concatenating multiple workflows (line 29). 

We make some simplifying assumptions in configuring the transition matrix to avoid combinatorial explosion and make the transitions tractable. However, we keep this transition matrix configurable to generate a diverse variety of workloads. The dependency of an interaction on the data scope (line 23) is implemented by sampling the system operation also considering the data attributes in the data scope. 
Unlike prior graph benchmarks which have a fixed set of operations to evaluate the test systems on, our benchmark creates INVA workloads considering the supported interactions common to all the test systems which are being evaluated ($[t_s]$) (line 23). This enables a fair comparison when creating workloads to evaluate test systems supporting different sets of interactions.

\renewcommand{\algorithmicrequire}{\textbf{Input:}}
\renewcommand{\algorithmicensure}{\textbf{Output:}}

\begin{algorithm}
\caption{Workload Generation Pseudocode}
\label{alg:workload-generation}
\begin{algorithmic}[1]

\REQUIRE 

\begin{enumerate}[leftmargin=*]
    \item Network Data ($D(N, L)$)
    \item Transition Probability Matrix ($M$)
    \item Workload length ($m$)
    \item Test systems ($[t_s]$)
\end{enumerate}

\ENSURE Workload $W = \langle w_1, w_2, \ldots, w_m \rangle$

\WHILE{$W.\text{length} < m$}

    \STATE // setting the data scope $D_j$ for workflow $w_j$
    \STATE $ D_s \gets \text{WeightedSample}(\{D, D_{j-1}\}) $
    \STATE $ \sigma \sim \mathcal{U}([\text{node\_filter, link\_filter, k-hop\_filter, none}])$
    \STATE $ dp(D_j) \gets \sigma(D_s)$
    \STATE $ da(D_j) \sim \mathcal{U}(Na \cup La, 2)$
    \STATE
    
    \IF{$ |N_j| = 0 $}
        \STATE \textbf{continue} ~~~~// reject empty data scope
    \ENDIF
    \STATE 
    
    \STATE // generating interaction sequence $I_j$ for workflow $w_j$
    \STATE $ I_j \gets \langle \rangle$
    \STATE $ k \gets 0$ ~~~~// interaction\_counter
    \WHILE{true}
        \STATE $i_k.af \gets M(i_{k-1}.af, k)$
        \STATE 
        
        \IF{$i_k.af = \text{'EOW'}$}
            \STATE \textbf{break} ~~~~// encountered 'end of workflow' token
        \ENDIF
        \STATE
        
        \STATE $i_k.at \gets M(i_{k-1}.af, i_k.af)$
        \STATE $i_k.op \gets \text{System\_Operations}(t_s, i_k.af, i_k.at, da(D_j))$
        \STATE $I_j.\text{add}(i_k)$
        \STATE $k \gets k + 1$
        
    \ENDWHILE
    \STATE 
    \STATE $w_j \gets (D_j, I_j)$
    \STATE $W.\text{add}(w_j)$ ~~~~// concatenating workflows to create workload

\ENDWHILE

\RETURN $W$

\end{algorithmic}

\end{algorithm}

We implemented the workload generator in Python using NetworKit library~\cite{staudt_networkit_2015} for \textit{graph} focus interactions and Dask library~\cite{dask} for \textit{attribute} and \textit{comparison} foci interactions.

\subsubsection{Representativeness of the INVA Model}

We studied prior work employing INVA workflows~\cite{chen_understanding_2020, ocallaghan_uncovering_2013, brandon_graph_2020, bodin_role_2009, wojcik_social_2021, charitou_biological_2016, wang_data-driven_2021, azizi_visibility_2025, kuikka_network_2025, hadaj_plans_2022, borsboom_network_2021, kalhor_insight_2022} to create our INVA model which enables the creation of workloads representative of how human analysts analyze network data. 

Analysts analyze the data with a certain analysis goal, and they begin by choosing a subset of the data or attributes to analyze, \ie \textbf{setting a data scope} (\cref{fig:inva-model} yellow box). This is followed by a sequence of interactions performed on the data scope as per the analysis goal (\cref{fig:inva-model} green box). Each interaction is conceived in terms of what property of the network is to be analyzed - topology, attributes or comparison between the two, \ie the \textbf{analysis focus}. The idea of \textbf{analysis task} gives more details about the kind of analysis to be performed for the interaction. Finally, the \textbf{system operation} provides an executable means to the end. We thus compose one workflow (\cref{fig:inva-model} blue box). Analysts may repeat the entire procedure multiple times with different analysis goals. Our model enables simulating this by generating multiple workflows and concatenating them to get a workload (\cref{fig:inva-model} gray box).
We keep the transition probability matrix configurable to enable modeling different INVA behaviors, but also provide a default transition matrix based on our observations from prior work employing INVA workflows. 
Our INVA model can also accommodate new interactions in a semantic way as per their focus and task category.

Our approach of generating INVA workflows (data scope + interaction sequence) is also corroborated by the kind of analysis constructs, certain tools like the Neo4j Graph Data Science~\cite{neo4jgds} plugin facilitate. \Cref{sec:improving-benchmark} describes future work to validate and further improve the workload generator. In its current state however, our generator can be used to generate workloads so that INVA systems can be tested for preliminary performance results, before evaluating them more thoroughly with human network analysts. These preliminary results can then be used to guide the research and development of new systems for INVA. Our model is not meant to substitute human network analysts, but rather to generate representative INVA workloads in order to test graph systems.



\subsection{Data Generator Module}
\label{sec:data-generator-module}

Prior works in network data generation either do not support node/link attribute generation~\cite{iosup_ldbc_2020, erling_ldbc_2015}, or support limited connectivity patterns or require expensive hardware and infrastructure~\cite{darabi_framework_2025, razavi_hierarchical_2026, alam_fast_2021}. We address these limitations in our own implementation of the network data generator module.

Our data generator (\cref{fig:benchmark-architecture}b) enables creating synthetic network datasets using a specification for the node and link attributes, the distribution of these attribute values, and the connectivity properties of the network---degree distribution, link probability and node:link ratio. Workloads can be created, and INVA systems can be evaluated on these synthetic datasets as an alternative to real-world datasets. We implemented the data generator in Python using the NetworKit network analysis library~\cite{staudt_networkit_2015}.



\subsection{Benchmark Driver}
\label{sec:benchmark-driver}

The benchmark driver module (\cref{fig:benchmark-architecture}c) serves as the interface between the benchmarking framework and the test system. It serves three purposes (1) provides a blueprint for test system interface development, (2) scaffolds running the workload on the test system, and (3) logs the results for offline analysis. We refer our readers to the supplementary material for more details about this module. We implemented the benchmark driver module in Python.






\section{Evaluation}
\label{sec:evaluation}

In this section, we describe the evaluation metrics, the setup used to demonstrate our benchmarking framework, and present benchmarking results for the \textbf{backends} of three types of graph systems as per the categorization presented in~\cite{sinva_survey_2026} - INVA systems, graph databases (GDB) and graph scripting libraries (GSL).

Our main focus is on evaluating graph systems for INVA use cases which are more human-oriented. Nevertheless, the underlying system performance is also crucial. Thus, we consider two types of metrics: Performance metrics, and Expressiveness metrics. We describe the performance metrics, which are more traditional or system-oriented in the following subsection, and expressiveness metrics which are more human user or developer oriented, along with the corresponding results in~\cref{sec:expressiveness-metrics-results}.

\subsection{Performance Metrics}
\label{sec:performance-metrics}

\subsubsection{Scalability Limit}
The upper limit on dataset size beyond which the test system fails to execute the workload. The failure could be due to any reason~\eg out-of-memory (OOM), no response for a long time (Timeout), failure to load the data correctly,~\etc


\subsubsection{Data Load Time}
The time to ingest the data in the system.

\subsubsection{Response Time for Interactions}
The time between issuing an interaction and getting a response. This is measured for both data scope operations and interaction system operations.

\subsubsection{Correctness}
Although most graph operations are guaranteed to return a deterministic and exact result across different graph systems, some graph operations are implemented using variants or approximations in different graph systems~\eg betweenness and closeness centrality. To enable a fair comparison of the correctness of such graph operations across systems, we use a multi-pronged approach for this metric, partly borrowing some approaches from the LDBC Graphalytics Benchmark~\cite{iosup_ldbc_2020}:


\begin{enumerate}[wide=2mm, leftmargin=4mm, itemsep=1mm]

\item[\textbf{Exact}:] For deterministic queries which return exact answers like data scope size, number of connected components, shortest path \etc we report their \textbf{accuracy} by comparing the exact values (exact match~\cite{iosup_ldbc_2020}) of test system results, with those of the workload generator (reference results). \label{eval-metric-exact}


\item[\textbf{Ranking}:] For deterministic queries, the results for which differ slightly owing to algorithmic variants,~\eg centrality measures, we report \textbf{accuracy} by computing the Spearman's ranking correlation coefficient between rankings of the nodes/links as per the computed attribute (for a limited sample of nodes/links) for the test system and the workload generator (reference results). \label{eval-metric-ranking}





\end{enumerate}

We do not report correctness for graph operations which are stochastic in nature or may be configured by the user to return different results \eg graph clustering and layout generation, due to the lack of a single correct answer for the same.

\subsubsection{Workload Completion Time}
The time to run the entire workload \textbf{as experienced by the human analyst}. This includes the time taken to load data, set data scopes and execute system operations for all workflows in the workload, in addition to the transformations required between intermediate data formats, from fetching query results to displaying the results on screen during the execution of workloads. It does not include the think time required by the human analyst to understand the results.

\subsection{Experimental Setup}

\subsubsection{Hardware Configuration}

We performed our experiments on a computer with a 48 core Intel Xeon 2.6 GHz CPU, having an L1 cache of 768 KB per core, L2 cache of 6 MB per core, and a shared L3 cache of 60 MB, 512 GB of main memory and 3.2 TB of disk space, running Rocky Linux 9.6. For reproducibility, better portability and sandboxing of the setup, 
we dockerized the benchmarking framework along with the test systems~\footnote{Docker container available at - \href{https://osf.io/3z4hk/overview?view_only=21e3de8aea404b1186c4d02938c24996}{https://osf.io/inva-benchmark}}. The docker container was configured to use Ubuntu 22.04 with 128 GB of main memory, and an additional 128 GB of swap space.

\subsubsection{Test System Selection}
\label{sec:test-system-selection}

We evaluate one INVA system-- Cytoscape~\cite{shannon_cytoscape_2003} and two GDBs -- Neo4j~\cite{neo4j} and Memgraph~\cite{memgraph}. We also report results on Tulip~\cite{auber_tulip_2004} (INVA system) for data load time and all expressiveness metrics, but we do not run workloads on Tulip (no response time and correctness metrics) due to difficulties in implementing the Tulip interface owing to insufficient documentation for the APIs exposed for its capabilities. We consider the workload generator results generated using Python NetworKit~\cite{staudt_networkit_2015}, a graph scripting library (GSL) as reference results. 

Apart from the aforementioned chosen test systems, we also considered Gephi~\cite{bastian_gephi_2009} (INVA system), Python NetworkX~\cite{hagberg_exploring_2008} (GSL), TigerGraph~\cite{tigergraph} and Kuzu~\cite{kuzu_2023} (GDB). We rejected Gephi due to insufficient resources to develop the interface to the benchmark driver. We rejected NetworkX due to known scalability issues, and Kuzu because it does not have its own graph analytics capabilities (it uses NetworkX instead, which limits its scalability). We were able to successfully support TigerGraph and evaluate it using our benchmark but cannot report benchmarking results to remain in legal compliance.

We thus evaluate representatives from three categories of graph systems - INVA, GDB and GSL, which provides a reference benchmark performance for systems in these categories on INVA workloads, and helps us understand how different are they from each other. Our implementation challenges also demonstrate why a benchmark is critical for INVA use cases; otherwise, many of these tools will continue to lack basic interactive features, scalability tests, and benchmarking support.

\subsubsection{Test System Configuration}

We now describe specific configurations for each chosen test system and their interfaces, using which the experiments were run.

\noindent\ding{226}\textit{Cytoscape~\cite{shannon_cytoscape_2003} (Java):} We used Cytoscape v3.10.4 with its inbuilt CyREST API and the python driver package py4cytoscape~\cite{py4cytoscape} v1.12.0 to implement the interface. The use of py4cytoscape requires the Cytoscape GUI tool to be running alongside. We run our experiments on a headless machine, but because Cytoscape requires a display to run the GUI tool, we create a virtual display using Xvfb utility.

\noindent\ding{226}\textit{Tulip~\cite{auber_tulip_2004} (C++):} We used Tulip python package tulip-python~\cite{tulip_python} v6.0.0 to write the Tulip interface. Unlike Cytoscape, the tulip-python driver can be used standalone and does not require the GUI tool running alongside.

\noindent\ding{226}\textit{Neo4j~\cite{neo4j} (Java):} We used Neo4j v2025.08.0 Community Edition along with the python driver neo4j~\cite{neo4jpy} v5.28.0, and additional utility libraries designed for network analysis---Neo4j Graph Data Science Library~\cite{neo4jgds} v2.21.0 and Neo4j APOC library~\cite{neo4japoc} v2025.08.0 to implement the interface. Neo4j also provides a visualization library - Neo4j Bloom. However, we do not use it in our experiments because it is only available for the enterprise edition of Neo4j. The Neo4j server is configured to use a maximum of 127 GB of heap memory. Additional server configuration settings can be found in the neo4j.conf file in the docker container.





\noindent\ding{226}\textit{Memgraph~\cite{memgraph} (C++):} We used Memgraph v3.7.2, an in-memory graph database, along with its inbuilt network analysis library--Memgraph Advanced Graph Extensions (MAGE)~\cite{mage}. Memgraph uses the same python driver package as Neo4j, with custom tweaks made for Memgraph. We were thus able to use Claude AI to easily adapt the Neo4j interface for Memgraph. The Memgraph server is configured to enable optimization for graph OLAP as opposed to graph OLTP, with a maximum query execution time of 3 hours after which it times out. Additional server configuration settings can be found in the memgraph.conf file in the docker container.

\noindent\ding{226}\textit{NetworKit~\cite{staudt_networkit_2015} (C++):} We used NetworKit v11.1.post1.

Of the chosen test systems, Neo4j, Tulip and Memgraph
have multiple driver options available. 
However we use the Python driver to keep our codebase uniform, for ease of integration, and to enable fair comparison with other systems which have only Python drivers. 
In a way, we test not only the test system backends, but also the python interfaces to these backends for executing various interactions. Thus the results presented can also help developers and researchers to decide which interface to use when developing applications on top of graph systems.

For a fair comparison, we ran our experiments on the single machine execution setup of all the test systems, even for systems which support a distributed setup. We also unit tested each test system interface to prevent any interface implementation issues from confounding the results of our evaluation. Our drivers run the workloads against test system \textbf{backends}, hence the benchmarking results do not reflect the extra cost incurred by GUI interaction processing, but they do account for the time spent in processing the python driver API calls. Finally, for system operations like graph layout, graph clustering~\etc which have multiple algorithm options available, we evaluate the default algorithm set in the system/GUI. We essentially evaluate the system for users having limited experience with using the system - they would resort to the defaults.


\subsubsection{Datasets \& Workloads} We evaluated the test systems using 8 datasets. One workload was generated for each dataset. The dataset and workload characteristics are shown in~\cref{tab:evaluation_dataset_characteristics} and \cref{tab:evaluation_workload_characteristics} respectively. The synthetic datasets were generated using our data generator module, while the real-world datasets (highlighted in~\cref{tab:evaluation_dataset_characteristics}) were obtained from the Network Repository~\cite{rossi_network_2015}. All the network datasets used for testing are undirected. We do so to allow for benchmarking of as many system operations across as many test systems as possible; most algorithms for directed networks also work on undirected networks, but the reverse is not true. We run each workload on each test system \textbf{thrice} and present both the individual and averaged results.


\begin{table*}[!ht]
    \label{tab:evaluation_dataset_characteristics}
    \caption{Characteristics of the datasets used for evaluation, listed in increasing order of data scale (last column). Real-world datasets are highlighted in yellow, the rest are synthetic datasets generated using our data generator.}
    \begin{tabular}{ccccccc}
        \toprule
         \textbf{Dataset} & \textbf{\#Nodes} \scriptsize{$|N|$} & \textbf{\#Links} \scriptsize{$|L|$} & \textbf{Node Attributes} & \textbf{Link Attributes} & \textbf{$|N|:|L|$ ratio} & \textbf{Log Data Scale} \scriptsize{$log_{10}(|N|+|L|)$} \\
        
        \cmidrule(ll){1-1} \cmidrule(ll){2-7}

        D1 & 1,000 & 1020 & Float: 2, String: 2 & Float: 1, Bool: 1 & 1:1 & 3.305 \\

        D2 & 50,000 & 49,760 & Float: 2, Bool: 2 & Float: 1, String: 2 & 1:1 & 4.999\\

        D3 & 1,000 & 100,055 & Float: 2, String: 2 & Float: 1, Bool: 1 & 1:100 & 5.005 \\

        \rowcolor{Goldenrod!40} D4 & 203,769 & 234,355 & Int: 1, Float: 3 & None & 1:1.15 & 5.642 \\

        \rowcolor{Goldenrod!40} D5 & 822,942 & 1,348,374 & String: 1, Int: 1, Float: 2 & None & 1:1.6 & 6.337 \\

        D6 & 50,000 & 5,003,843 & Float: 2, Bool: 2 & Float: 1, String: 2 & 1:100 & 6.704 \\

        \rowcolor{Goldenrod!40} D7 & 2,146,057 & 5,743,132 & None & Int: 2 & 1:2.7 & 6.897 \\

        D8 & 600,000 & 15,003,456 & Float: 2 & Float: 2 & 1:25 & 7.193 \\
        
        \bottomrule
    \end{tabular}
\end{table*}

\newcolumntype{E}{>{\centering\arraybackslash}p{0.07\textwidth}}

\begin{table}[t]
    \label{tab:evaluation_workload_characteristics}
    \caption{Characteristics of the evaluation workloads. Each workload is created for the corresponding numbered dataset.}
    \begin{tabular}{EEEEE}
        \toprule
        
             \textbf{\rot{Workload (\#workflows)}} & \textbf{\rot{\#data scope operations}} & \textbf{\rot{\#graph focus interactions}} & \textbf{\rot{\#attribute focus interactions}} & \textbf{\rot{\#comparison focus interactions}} \\ 
        
        \cmidrule(ll){1-1} \cmidrule(ll){2-2} \cmidrule(ll){3-5}

            W1 (8) & 5 & 17 & 10 & 11 \\
    
            W2 (9) & 6 & 7 & 7 & 6 \\
    
            W3 (7) & 6 & 6 & 7 & 12 \\
    
            W4 (9) & 7 & 9 & 12 & 7 \\
    
            W5 (8) & 4 & 9 & 13 & 14 \\
            
            W6 (9) & 8 & 8 & 8 & 8 \\
    
            W7 (6) & 5 & 3 & 6 & 4 \\
    
            W8 (7) & 7 & 7 & 9 & 5 \\
    
        \bottomrule
    \end{tabular}
\end{table}

\subsection{Performance Results}

In this section, we report the performance results of the chosen test systems on our benchmark, along with the workload generator (NetworKit) results which we consider as reference results. 






\subsubsection{Scalability Limit}
As shown in~\cref{tab:scalability-limit-results}, Neo4j, Memgraph and NetworKit scaled up to the largest dataset \textit{D8} without issues. Cytoscape scaled up to \textit{D4} with timeout failures for \textit{D5}, \textit{D7} and \textit{D8}. We observe that \textbf{graph DBMSs (GDB) are generally designed for scalability more than INVA systems} as also observed in prior work~\cite{sinva_survey_2026}. The OpenMP based parallelized implementation of graph operations in NetworKit partially explains its scalability.


\begin{table}[t]
    \label{tab:scalability-limit-results}
    \caption{Scalability limit results; Log data scale reported in parentheses; \textit{Timeout} means no response for 3 hours.}
    \begin{tabular}{ccccc}
        \toprule
        \textbf{Dataset} & \textbf{Cytoscape} & \textbf{Neo4j} & \textbf{Memgraph} & \textbf{NetworKit}\\
        
        \cmidrule(ll){1-1} \cmidrule(ll){2-5}

        D1 (3.305) & \cmark & \cmark & \cmark & \cmark \\

        D2 (4.999) & \cmark & \cmark & \cmark & \cmark \\

        D3 (5.005) & \cmark & \cmark & \cmark & \cmark \\

        D4 (5.642) & \cmark & \cmark & \cmark & \cmark \\

        D5 (6.337) & \textit{Timeout} & \cmark & \cmark & \cmark \\

        D6 (6.704) & \cmark & \cmark & \cmark & \cmark \\

        D7 (6.897) & \textit{Timeout} & \cmark & \cmark & \cmark \\

        D8 (7.193) & \textit{Timeout} & \cmark & \cmark & \cmark \\
        
        \bottomrule
    \end{tabular}
\end{table}



\subsubsection{Data Load Time}
\begin{figure}
    \centering
    \includegraphics[width=\linewidth]{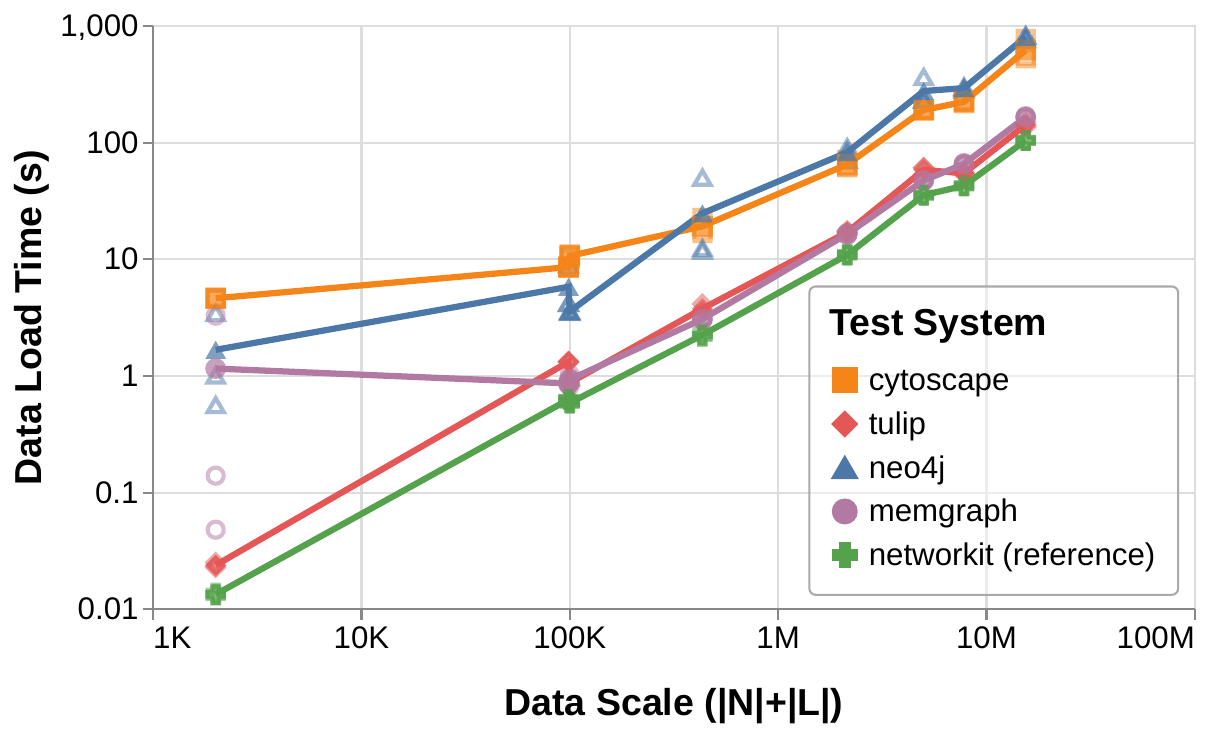}
    \caption{Average data load time across 3 runs with raw values \vs data scale for the test systems, along with workload generator (NetworKit) reference results.}
    \label{fig:data-load-time}
\end{figure}

\Cref{fig:data-load-time} shows that Cytoscape and Neo4j consistently take more time than Tulip, Memgraph and NetworKit, with the difference being around an order of magnitude at the 100K data scale mark and beyond. Both Cytoscape and Neo4j are implemented in Java and store nodes/links using linked lists. In contrast, Tulip, Memgraph and NetworKit are implemented in C++, and store nodes/links in contiguous memory arrays (vectors) which partially explains why they perform better than Cytoscape and Neo4j. NetworKit uses OpenMP for parallelized execution which also explains its faster speed especially at smaller data scales. Memgraph is an in-memory GDB as opposed to Neo4j which is a disk-based, which also explains their performance difference. Furthermore, the results for Neo4j are obtained with explicit indexing on nodes and batching during data load, without which Neo4j takes an unreasonably long time for loading data, and yet it is marginally slower than Cytoscape for more than the 1M data scale mark. In all, we see \textbf{clear performance difference between systems which use linked-list \vs contiguous memory based network data storage.}


\subsubsection{Data Scope Operation Response Time \& Correctness}
\begin{figure*}
    \centering
        \includegraphics[width=\linewidth]{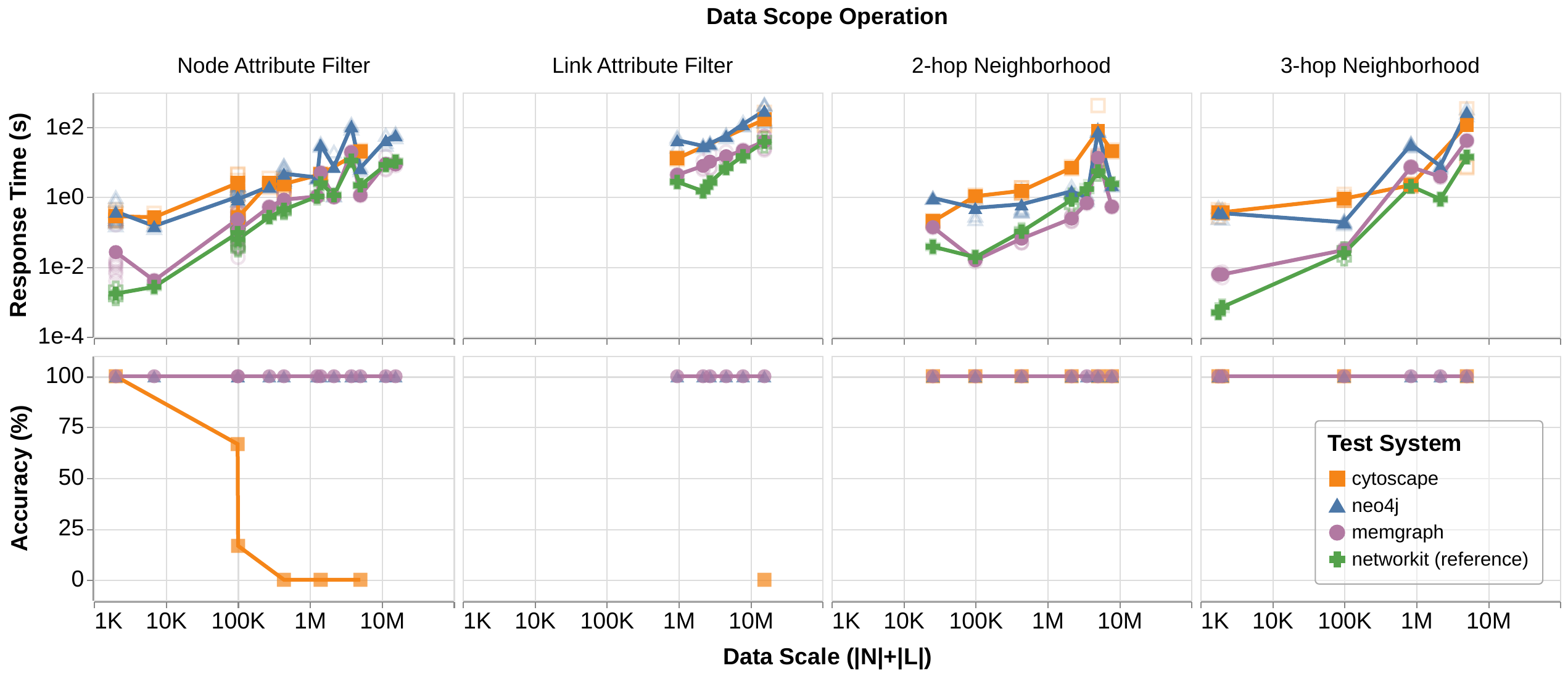}
        \caption{Average response time across 3 runs with raw values (top row) and average accuracy across 3 runs (bottom row) \vs data scale for the test systems. Accuracy plots show overplotting for test systems which report 100\% accuracy.} 
        \label{fig:data-scope-operation-time-accuracy}
    
\end{figure*}

\Cref{fig:data-scope-operation-time-accuracy} (bottom row) shows that all test systems except for Cytoscape perform data scope operations correctly at all scales. Although Cytoscape returns correct results at all scales for the k-hop neighborhood operations, it returns incorrect results for node and link attribute filters around the 100K data scale mark and beyond, with inconsistent reproducibility. \textbf{The Cytoscape team has acknowledged the intermittent correctness issue at scale, uncovered by our benchmark.}


In terms of response time, \cref{fig:data-scope-operation-time-accuracy}~(top row) shows that when Cytoscape does work, it performs almost comparably (within an an order of magnitude) with Neo4j across all data scope operations and data scales. However Cytoscape does not scale as much as Neo4j as shown both in~\cref{tab:scalability-limit-results} and~\cref{fig:data-scope-operation-time-accuracy}. Memgraph performs at least around an order of magnitude faster than Cytoscape and Neo4j for all data scope operations with minor variations \wrt data scale. The faster execution of Memgraph compared to Cytoscape and Neo4j could be attributed to the use of vectors in Memgraph \vs linked lists in Cytoscape and Neo4j. The in-memory nature of Memgraph further strengthens its case compared to Neo4j. Thus for data load time, \textbf{Cytoscape is among the slower test systems, followed by the GDBs, while surprisingly, NetworKit is the fastest across all data scope operations and nearly all data scales}, with exceptional performance for k-hop neighborhood operations.

\subsubsection{System Operation Response Time \& Correctness}
\begin{figure}
    \centering
    \includegraphics[width=\linewidth]{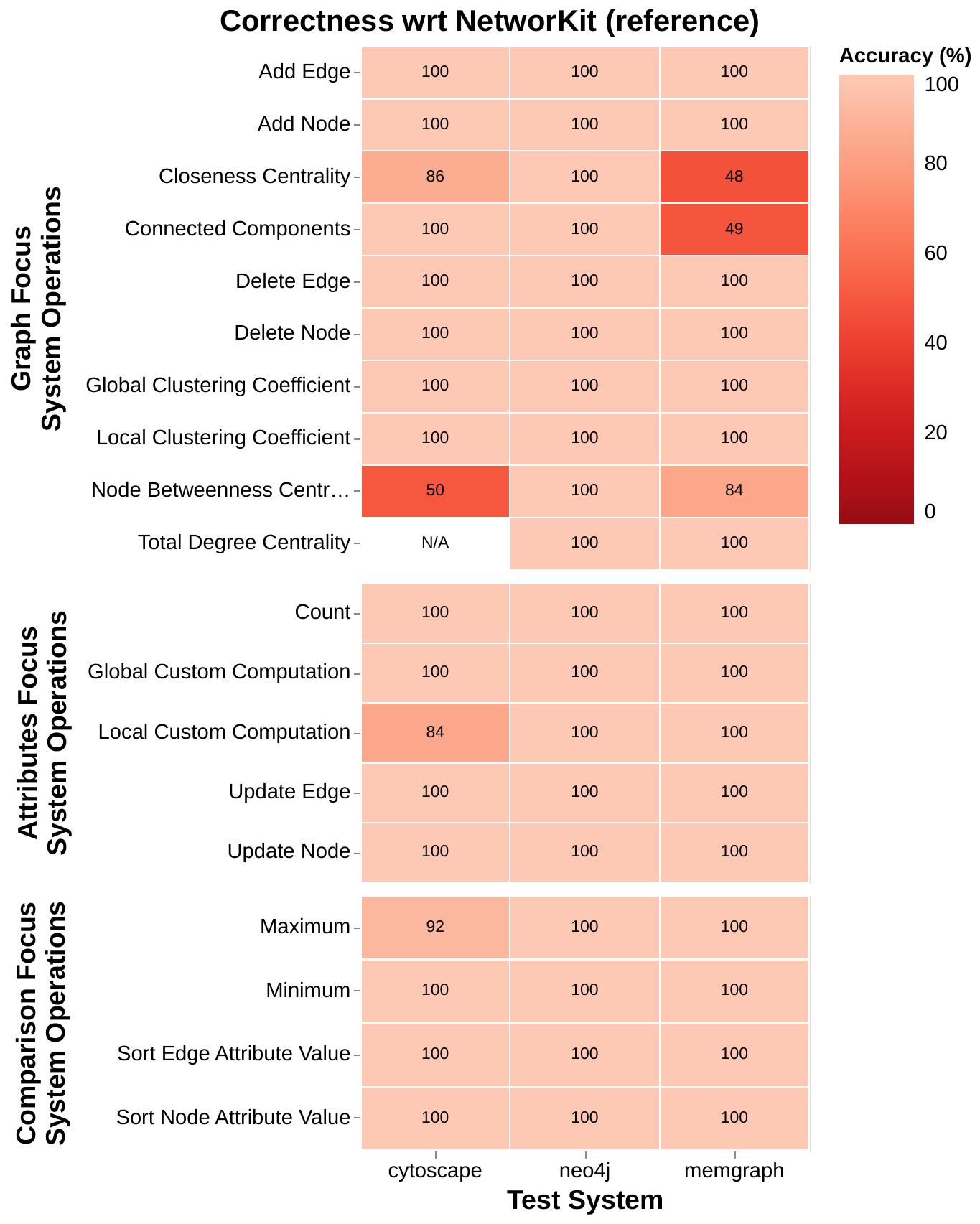}
    \caption{Summary of correctness in terms of accuracy, of evaluated system operations for different test systems \wrt NetworKit (reference) results, averaged across all data scales.}
    \label{fig:correctness-results-summary}
\end{figure}

To save space, we present results for only a select few system operations which reveal correctness differences across test systems due to differences in implementation, as shown in~\cref{fig:correctness-results-summary}. For a meaningful comparison, we evaluate correctness of a system operation for a given test system only when the test system is able to correctly execute the data scope operation of the corresponding workflow within the workload. As a reference for how responsive each system operation is for INVA use cases, we annotate all response time plots with a dashed red line at 1s denoting an interactivity threshold which when exceeded, can disrupt the flow of thought of the analyst~\cite{nielsen_response-times_1993}. Given the lack of prior work evaluating interactivity specifically for graph analysis, we call for future work in this space in~\cref{sec:olap-vs-oltp-for-graphs}.

\begin{figure}
    \centering
    \includegraphics[width=\linewidth]{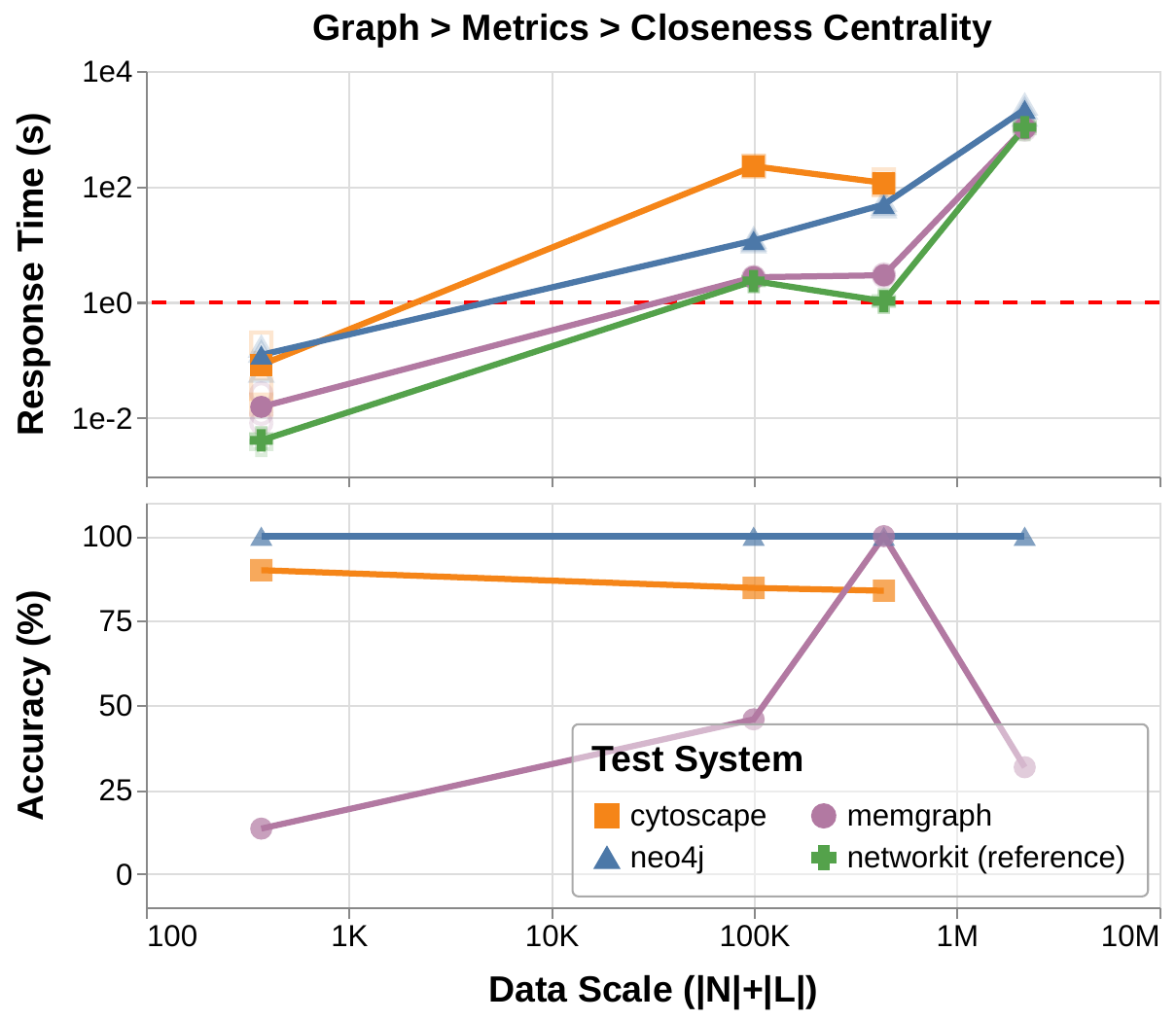}
    \caption{Results for reponse time and correctness (\hyperref[eval-metric-ranking]{\textcolor{teal}{\textit{ranking}}} based accuracy) for computing closeness centrality}
    \label{fig:closeness-centrality-results}
\end{figure}

\textit{Closeness Centrality.} 
For the limited number of instances of \textit{closeness centrality} system operation, \cref{fig:closeness-centrality-results} shows that both Memgraph and NetworKit consistently run around an order of magnitude faster than Cytoscape and Neo4j. While Neo4j runs faster than Cytoscape, NetworKit runs faster than Memgraph. 
All test systems fail to meet the 1s interactivity threshold once the data scale reaches the 100K mark.
Although Memgraph is one of the faster graph systems evaluated in our experiments, it has severe correctness issues with accuracy dropping down to less then 50\% in some cases. The centrality values for all failed instances of Memgraph agree with the reference results except for some nodes for which the closeness centrality value is reported as `undefined'. These are isolated nodes in the data scope to which Neo4j, Cytoscape and NetworKit assign a value of 0. Cytoscape also shows an accuracy between 80\% and 90\% across all instances, which possibly hints towards an approximated implementation of closeness centrality in Cytoscape. Thus \textbf{our benchmarking framework revealed possible differences in Cytoscape, and possible bugs in Memgraph for closeness centrality computation.} 

\begin{figure}
    \centering
    \includegraphics[width=\linewidth]{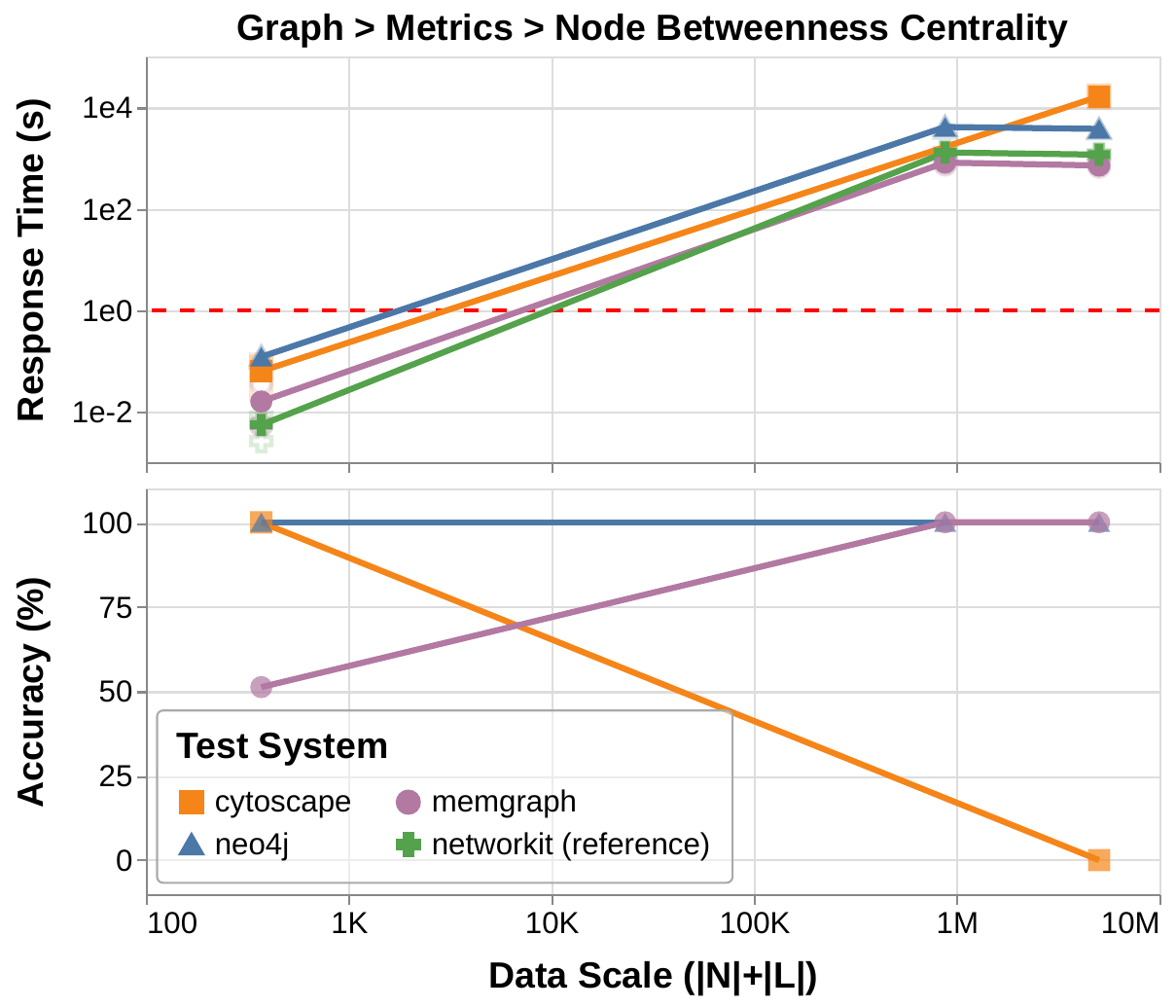}
    \caption{Results for response time and correctness (\hyperref[eval-metric-ranking]{\textcolor{teal}{\textit{ranking}}} based accuracy) for computing node betweenness centrality}
    \label{fig:node-betweenness-centrality-results}
\end{figure}

\textit{Node Betweenness Centrality}
For the limited number of instances of node betweenness centrality, \cref{fig:node-betweenness-centrality-results} shows that all 3 test systems along with NetworKit have relatively similar runtimes at lower scale compared to at higher scale, where Cytoscape might not be as fast as others. There is insufficient data to comment on at what data scale node betweenness centrality fails to meet the 1s interactivity threshold. Memgraph shows reduced correctness with the same issue as that for closeness centrality--unreachable nodes being assigned an undefined value instead of 0. With 0\% accuracy at close to the 10M data scale mark, \textbf{our benchmark reveals another possible bug in the node betweenness centrality computation at scale for Cytoscape.}

\begin{figure}
    \centering
    \includegraphics[width=\linewidth]{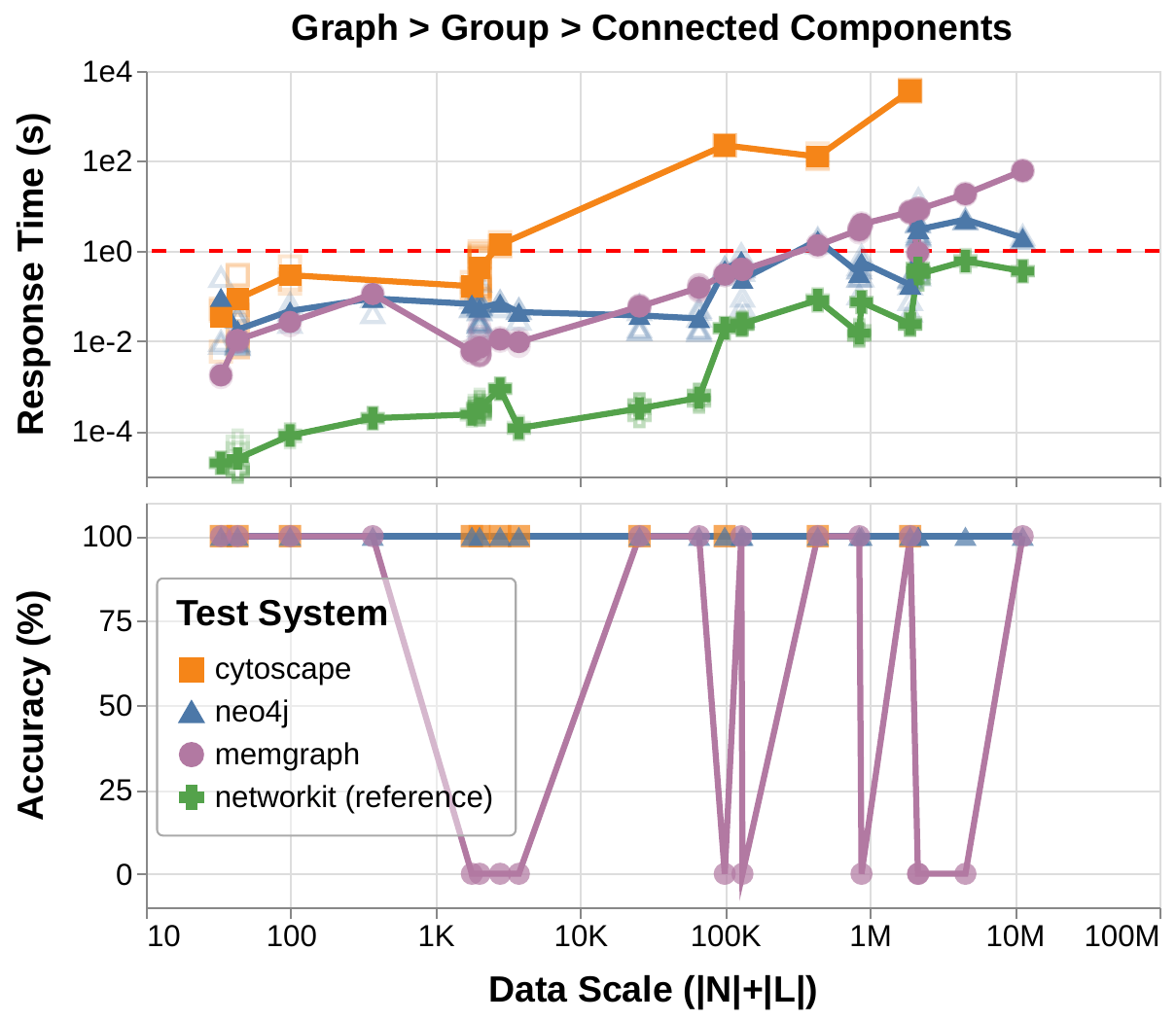}
    \caption{Results for response time and query correctness results (\hyperref[eval-metric-exact]{\textcolor{teal}{\textit{exact}}} accuracy) for counting the number of connected components}
    \label{fig:connected-components-results}
\end{figure}



\textit{Connected Components} As shown in \Cref{fig:connected-components-results}, Cytoscape runs around two orders of magnitude slower than the GDBs as the data scale increases. It also fails to meet the interactivity threshold before the 10K data scale mark. Memgraph and Neo4j have similar runtimes and remain interactive roughly up to the 1M data scale mark. While both Memgraph and Neo4j are around an order of magnitude slower than NetworKit, this difference decreases as the data scale increases. 
However in spite of its competitive runtime, \textbf{Memgraph has severe correctness issues in some cases suggesting a possible bug in its connected components implementation.}









\begin{figure}
    \centering
    \includegraphics[width=\linewidth]{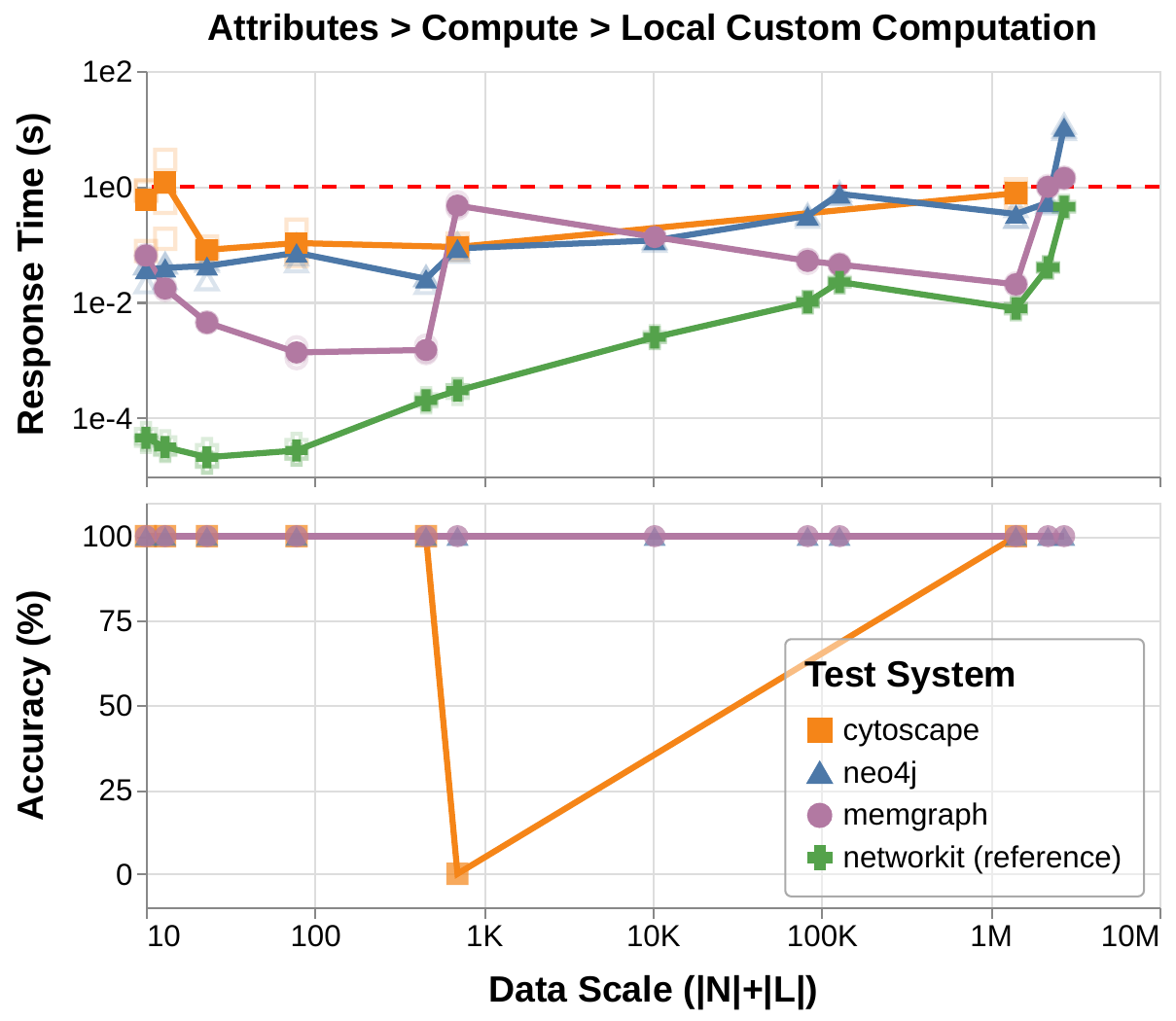}
    \caption{Results for response time and query correctness (\hyperref[eval-metric-exact]{\textcolor{teal}{\textit{exact}}} accuracy) for transforming a node/link attribute \ie local custom computation}
    \label{fig:local-custom-computation-results}
\end{figure}

\begin{figure}
    \centering
    \includegraphics[width=\linewidth]{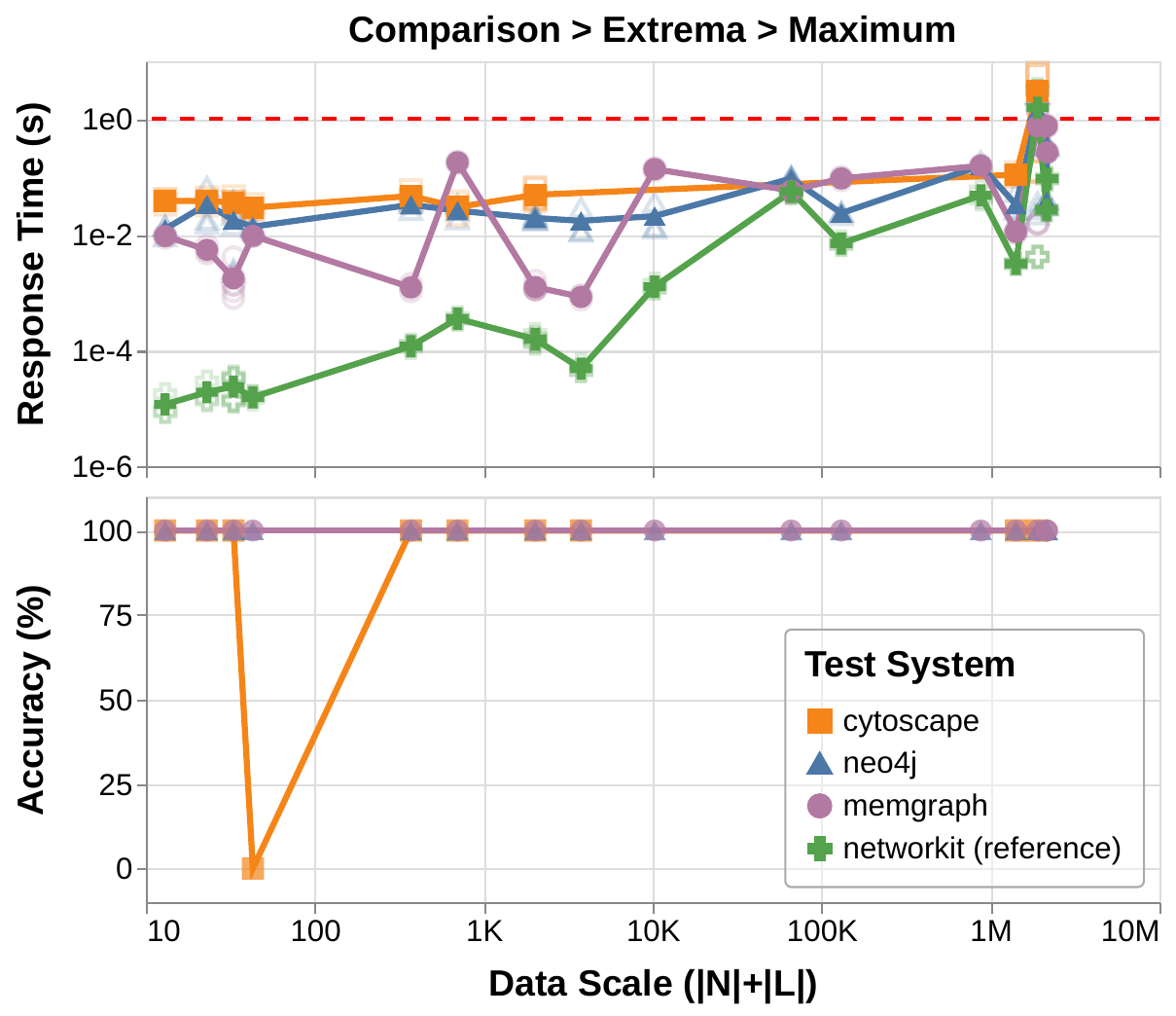}
    \caption{Results for response time and query correctness (\hyperref[eval-metric-exact]{\textcolor{teal}{\textit{exact}}} accuracy) for finding the maximum value of a node/link attribute}
    \label{fig:maximum-results}
\end{figure}

\textit{Local Custom Computation \& Maximum} \Cref{fig:local-custom-computation-results} and \cref{fig:maximum-results} show that the runtime for attribute and comparison focus system operations increases linearly \wrt data scale but with a smaller constant factor as opposed to graph focus system operations. This can be explained by the time complexity of attribute and comparison focus system operations which is either $O(|N|)$ or $O(|L|)$, whereas for graph focus system operations the time complexity is generally $O(|N|+|L|)$ (\eg connected components), $O(|N|.|L|)$ (\eg centrality measures) or worse. Because of this, these system operations can meet the interactivity threshold up to slightly beyond the 1M data scale mark, which is more than that for graph focus system operations. Comparatively, Cytoscape and Neo4j have similar runtimes while Memgraph runs faster in most cases followed by NetworKit which is the fastest. 
In terms of correctness, \textbf{Cytoscape has some issues for both \textit{local custom computation} and \textit{maximum} system operations, hinting towards possible bugs.}

\subsubsection{Workload Completion Time}
\begin{figure}
    \centering
    \includegraphics[width=\linewidth]{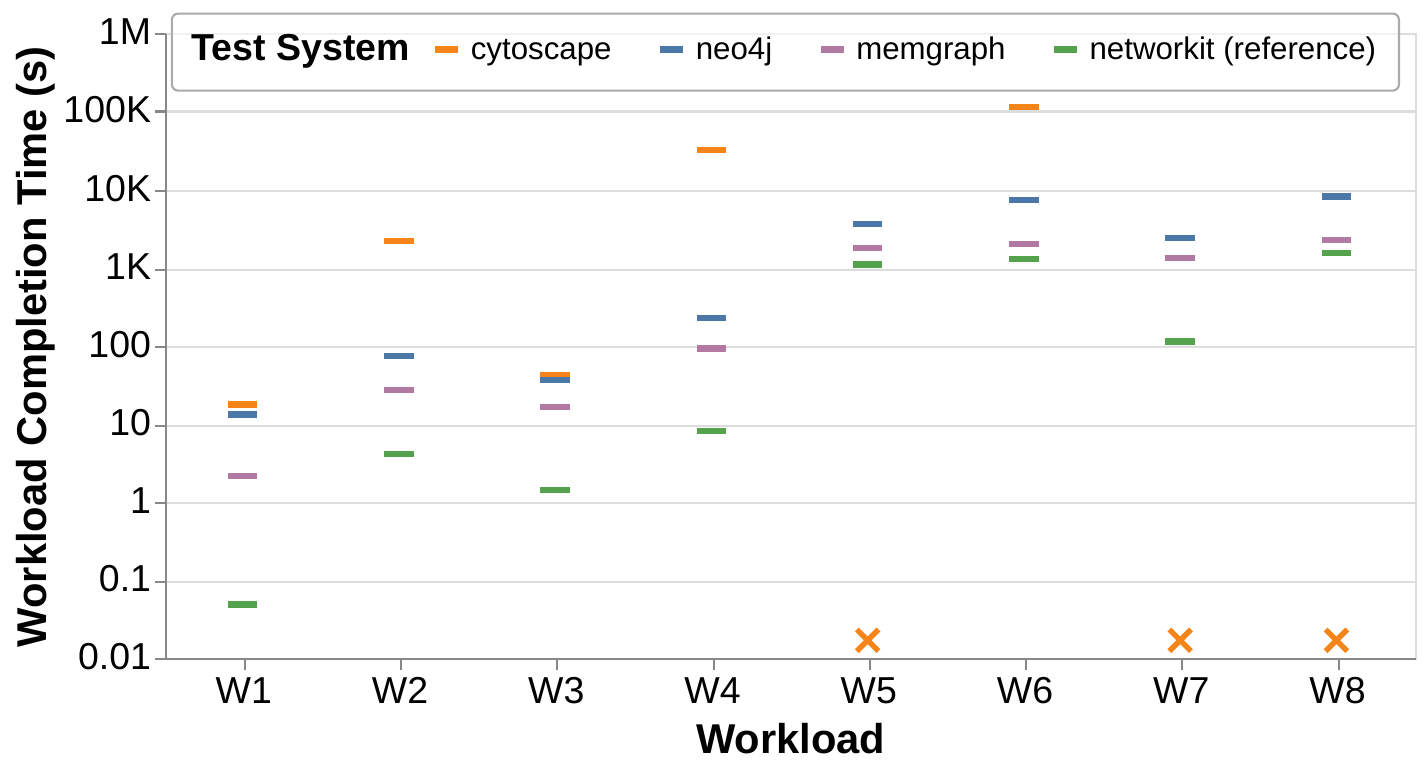}
    \caption{Workload completion time for each test system averaged across 3 runs. \ding{54} indicates failed workload run.}
    \label{fig:workload-completion-time}
\end{figure}

\Cref{fig:workload-completion-time} shows that, of the evaluated test systems, Memgraph is the most suitable for INVA in terms of scalability and response time. Although Cytoscape is a dedicated INVA system, it shows higher workload completion time along with scalability issues (W5, W7, W8). This shows how \textbf{graph databases are better equipped to be used for INVA for large networks than dedicated INVA systems.} While NetworKit is faster than all the test systems, its programming heavy nature can make it less favorable for domain experts or non-programmers.


\subsubsection{Takeaways}
Both in terms of correctness and response time, Cytoscape clearly struggled to compete with the GDBs, and NetworKit was comprehensively better than the GDBs. For all system operations across all three analysis foci, Cytoscape almost always failed to meet the interactivity threshold earlier \ie at a smaller data scale, than other test systems. Neo4j, Memgraph and NetworKit in that specific order, remain interactive up to a higher data scale mark than the prior system. 




While Neo4j performed with 100\% accuracy for all system operations across all data scales, our benchmarking evaluation showed that Memgraph has correctness issues. \textbf{The correctness bugs uncovered in Cytoscape and Memgraph demonstrate the value of our benchmark in performance and correctness testing for INVA workloads.} Cytoscape has acknowledged some of the bugs while Memgraph is yet to confirm the possible implementation bugs revealed by our benchmark. Going ahead, we need more experiments with different graph systems which support INVA, and at higher data scales, to understand if dedicated INVA systems like Cytoscape can still survive, and if GSLs like NetworKit can still remain competitive with GDBs.


\subsection{Expressiveness Metrics \& Results} 
\label{sec:expressiveness-metrics-results}

We now describe the more human user or developer oriented \textbf{soft} metrics, along with the results on our chosen test systems. These metrics help us gauge how easy or difficult it can be for analysts to use these systems, and for developers to work with them.

\subsubsection{Analyses Coverage} Percentage of system operations from the benchmark which are supported in the test system. 

\Cref{fig:analyses-coverage} shows that Cytoscape supports fewer graph focus system operations than Neo4j and Memgraph, \eg path finding, graph clustering~\etc and also does not facilitate a graph querying language. On the other hand, both Neo4j and Memgraph provide better support for graph focus interactions through their dedicated graph analysis libraries---Neo4j GDS and MAGE respectively. Cytoscape supports attribute focus system operations through node/link table manipulations in the GUI, while the same can be achieved using the Cypher query language facilitated in GDBs. Finally, Cytoscape exposes more capabilities to visualize the network data along with auxiliary visualizations for node/link attributes (comparison focus), than Neo4j and Memgraph. We had the limitation of working without Bloom visualization plugin for Neo4j Community edition, and Memgraph does not expose a visualization API.

\subsubsection{Lines of Code} Number of lines of code required to implement the test system interface using their respective driver APIs.

Both Neo4j and Memgraph require writing Cypher queries for data scope operations and some graph and attribute focus interactions. Furthermore, they allow the creation of only directed graphs, requiring explicit conversion from directed to undirected for certain interactions. These factors contribute to a higher Lines of Code compared to Cytoscape which exposes more of the required functionality through concise APIs as shown in~\cref{fig:loc-eoi}.

\subsubsection{Ease of Implementation} 5-point Likert scale rating based on the ease of availability and comprehensiveness of the documentation, developer support, and activity/responses on user forums, to help implement the test system interfaces to the benchmarking framework using the test system driver APIs. 



\begin{figure}
    \centering    
    \begin{subfigure}{\linewidth}
        \includegraphics[width=\linewidth]{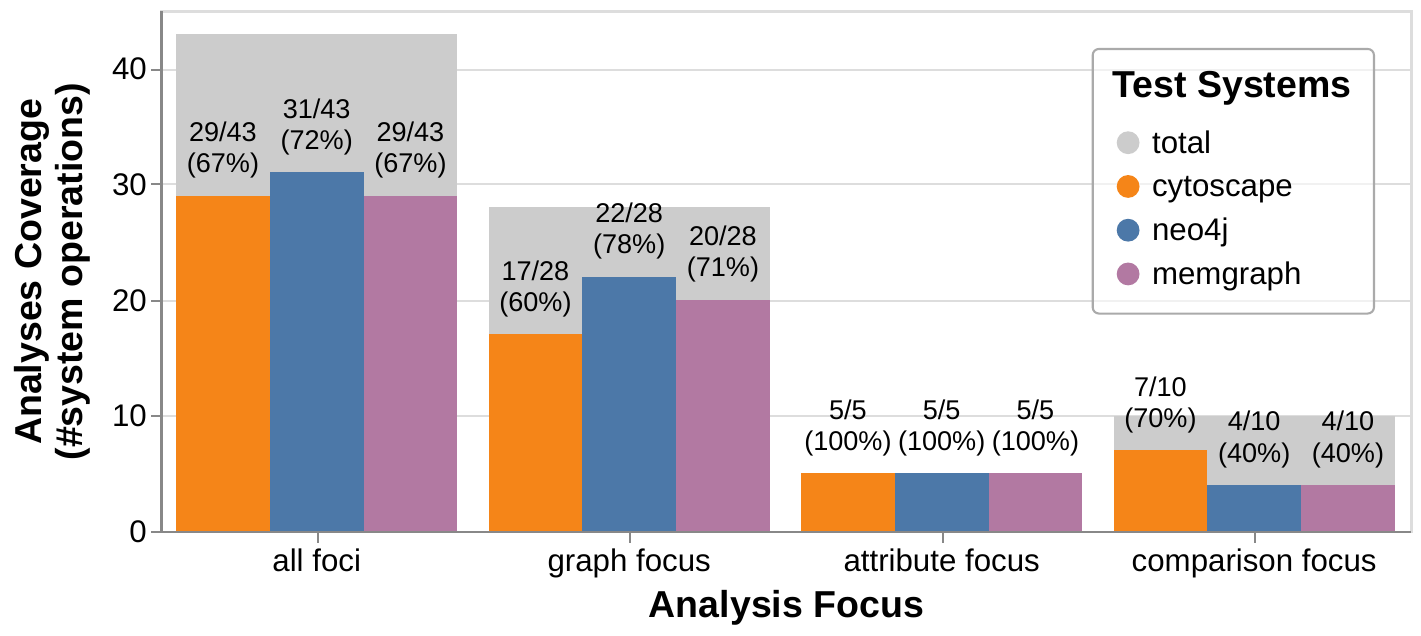}
        \caption{Analyses Coverage}
        \label{fig:analyses-coverage}
    \end{subfigure}
    \begin{subfigure}{\linewidth}
        \includegraphics[width=\linewidth]{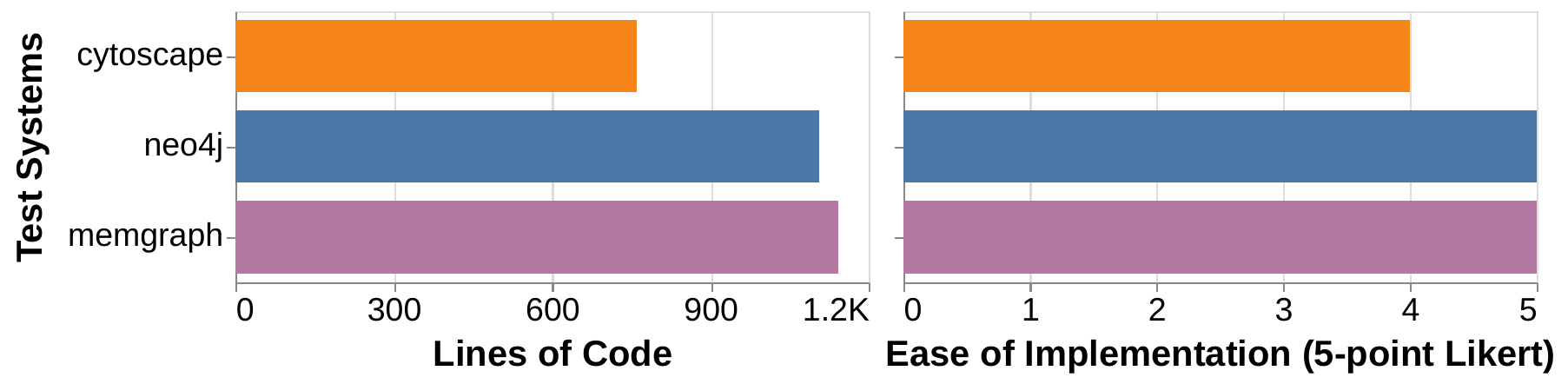}
        \caption{Lines of Code and Ease of Implementation}
        \label{fig:loc-eoi}
    \end{subfigure}
    \caption{Expressiveness metrics for the test systems}
    \label{fig:expressiveness-metrics}
\end{figure}


All three test systems---Cytoscape, Neo4j and Memgraph provide good developer support with active user base and user/developer forums. Owing to its recency, Memgraph does not have as large a user base as Neo4j, based on the number of followers for their respective github repositories. Neo4j and Memgraph both use the same python driver which is well documented, fetching them 5 points for Ease of Implementation. Cytoscape provides good documentation for the py4cytoscape API but not for CyREST API, and thus we gave it 4 points on Ease of Implementation as shown in~\cref{fig:loc-eoi}.

\section{Discussion}

In this section, we discuss some limitations and future work, and also highlight certain insights from our benchmarking evaluation.

\subsection{Towards Efficient Evaluation}
\label{sec:improving-benchmark}

Our INVA model does not cover exhaustively all possible system operations, and thus the generated workloads may not cover exhaustively all possible network analysis scenarios. However, our model facilitates adding more such system operations, and a configurable transition probability matrix, using which a variety of semantic INVA workloads may be generated. User analysis goal based transition probability matrices may also be used for more realistic workload generation~\cite{purich_adaptive_2025}. We also need formal validation of how closely the workloads generated using our model resemble the workflow of human network analysts. Since INVA emphasizes human-in-the-loop network data understanding, we also need user studies to understand the user experience with different graph systems for INVA. Such studies can compliment our evaluation of graph system backends thus providing a holistic picture of how good a graph system is for INVA. Based on our experience developing the interface for each test system, and the growing importance of graph applications across domains, we envision a unified interface for all graph systems, which could ease the benchmarking of different graph systems, also making it easy for developers to test different graph system backends when developing graph applications.

\subsection{INVA \vs Graph DBMS \vs Scripting Libraries}
\label{sec:inva-gdb-gsl}

The results of our benchmarking evaluation reveal certain insights which could be extended to the graph system categories of INVA systems, graph DBMS (GDB) and graph scripting libraries (GSL). In addition to providing network data storage solutions and efficient graph analytics, GDBs today also provide plugins for graph analysis and visualization accessible through GUIs, along with AI agents \eg Memgraph and FalkorDB~\cite{falkordb}, thus making them as convenient to use if not more, as INVA systems. This could be crucial for domain experts or non-programmers who form a significant portion of INVA systems user base. Newer GSLs are also outperforming dedicated INVA systems, even competing with GDBs up to the 1M data scale mark. This partially explains why newer GDBs are leveraging GSLs to provide graph analytics capabilities~\eg NetworkX for Memgraph and Kuzu. The only drawback of GSLs is the lack of a GUI which might restrict their user base to programming experts. With the inability of INVA systems to keep up with GDBs for scalability and interactivity requirements of INVA, and the increasing adoption of user friendly affordances in GDBs, GDBs seem to be well equipped to replace dedicated INVA systems.

\subsection{OLAP \vs OLTP for Graphs}
\label{sec:olap-vs-oltp-for-graphs}

The data management community interprets interactivity differently for graph OLTP and OLAP. Graph OLTP queries, which consist of queries operating on small sub-graphs, local neighborhoods or shortest path queries are considered interactive, but graph OLAP queries which consist of long running iterative algorithmic graph analysis~\eg centrality measures, community detection~\etc are not considered interactive~\cite{tian_world_2023, erling_ldbc_2015}. However Battle~\etal~\cite{battle_role_2020} have shown that the threshold for interactivity in visual analytics depends on the complexity of the task being performed, and with the current state of advancements in graph analysis hardware \eg GPU, and algorithms \eg parallelized algorithms, progressive algorithms, some of these tasks can be performed reasonably interactively. This raises an important question - \textit{\textbf{What is a reasonable interactive latency for INVA tasks?}} We thus need to rethink how we approach interactivity in graph OLAP.

 
\section{Related Work}
\label{sec:related-work}

\subsection{Benchmarking Graph Systems}
\label{sec:bg-benchmarking} 

Prior work in benchmarking graph systems, most of which comes from the Linked Data Benchmark Council (LDBC), is focused more on offline graph processing systems like Giraph, PowerGraph and GraphX~\cite{iosup_ldbc_2020, meng_revisiting_2025, dhulipala_graph_2020} or on interactive (graph OLTP) workloads but for graph databases~\cite{erling_ldbc_2015, armstrong_linkbench_2013} and not for INVA, which is the main focus of our work. Ours is the first benchmark to accommodate INVA workloads which iterate between interactive graph operations like inserts/deletes, operations on local neighborhoods~\etc and non-interactive graph operations like community detection, centrality measure computation~\etc
While the LDBC has separate benchmarks for different data domains like social network analysis~\cite{erling_ldbc_2015} and financial network analysis~\cite{qi_ldbc-financial_2025}, our benchmark can generate a variety of workloads using a configurable transition probability matrix, and for any network data domain, thus making it versatile. The benchmarking work of Eichmann~\etal~\cite{eichmann_idebench_2020} and Battle~\etal~\cite{battle_database_2020} for interactive analysis of relational data is more relevant to our work owing to its user behavior modeling in the interactive environment. Nevertheless, we refer to the work of LDBC for choosing datasets with appropriate diversity and coverage, and performance metrics for evaluation.

\subsection{Graph Task Taxonomies}
\label{sec:bg-graph-task-taxonomies}

We primarily base our INVA model on the graph task taxonomies of Lee~\etal~\cite{lee_task_2006}, Nobre~\etal~\cite{nobre_state_2019} and the operational graph task survey of~\cite{sinva_survey_2026}. Apart from that, Filipov~\etal~\cite{filipov_are_2023} provide a comprehensive survey of the prior work in graph task taxonomies, some of which we mention here. 
Pretorius~\etal~\cite{pretorius_tasks_2014} presented a more refined version of Lee~\etal's taxonomy. Kerracher~\etal~\cite{kerracher_task_2015} extended the same taxonomy for time-varying graphs, wherein they added an orthogonal dimension of time as timestamps or time intervals. The taxonomy of Ahn~\etal~\cite{jae-wook_ahn_task_2014} for dynamic graphs categorizes tasks along three aspects of network data - the type of data points being analyzed (nodes, links, clusters), the temporal aspect, and the data attributes. Saket~\etal~\cite{saket_group-level_2014} provide an alternative perspective on graph task taxonomies by considering clusters or groups of nodes instead of just raw nodes/links.

\section{Conclusion}

Acknowledging the growing need for human-in-the-loop or interactive understanding of large scale network data, we developed a benchmarking framework grounded in an empirical model of network visualization and analysis workflows, to comparatively evaluate different graph systems for interactive network visualization and analysis (INVA). We demonstrated the value of our benchmarking framework using one INVA system and two graph DBMSs. In addition to uncovering correctness issues, our findings show how purpose built INVA systems face the danger of being replaced by graph DBMSs given their inability to scale up to the large scale graphs in use today. We also call for consideration of interactivity in graph analysis so as to enable the development of scalable \textbf{and} user friendly systems to facilitate interactive network visualization and analysis.

\begin{acks}
    We thank the members of the IDL and Database labs at UW for their feedback, Alex Pico and team (Cytoscape), Xuanlei Lin and Songting Chen (TigerGraph), and Ante Javor (Memgraph) for helping with the development of the respective interfaces for use with our benchmarking framework, and explaining some of the results from our evaluation. This research was supported in part by Google and NSF awards IIS-2402718, IIS-2141506 and IIS-2514565.
\end{acks}


\bibliographystyle{ACM-Reference-Format}
\bibliography{sample}

\pagebreak
\appendix

\section{Benchmark Driver}
\label{sec:apx-benchmark-driver}

The benchmark driver module serves three purposes (1) provides a blueprint for test system interface development, (2) scaffolds running the workload on the test system, and (3) logs the results for offline analysis. In this section, we provide more details on how it serves as a blueprint for test system interface development.

The benchmark driver module has a base class which is to be derived by individual test system interface implementations to interface with the benchmarking framework. This base class defines the APIs for different operations for data scope selection and graph analysis. By default the base class implementation marks various operation implementations as `\textit{unsupported}'. Individual test system interfaces can override these operation methods in their respective derived test system interfaces, if they support that operation.



Different test systems may support a particular graph or data scope operation differently. Some test system drivers expose a single API to execute the desired operation, while some test system drivers may require using multiple APIs together to achieve the functionality, which may also go to the extent of writing our own implementation from scratch. Since we intend to benchmark the test system, and not our own algorithmic implementations, we use the following rubric to determine if we should implement the API manually for a graph or data scope operation for a given test system.

\begin{enumerate}[leftmargin=5mm, itemsep=3pt]
    \item Does the test system driver expose a single API or a set of APIs to be used together to execute the operation?
    If yes, use it/them. If no, continue to 2.

    \item Does the test system provide declarative constructs (\eg query language support) to help implement the interaction?
    If yes, continue to 3. If no, do not implement the interaction.

    \item Does the implementation around the available constructs require iterating over the data scope points manually?
    If yes, do not implement the interaction. If no, use the constructs and implement the interaction.
\end{enumerate}

\section{Ease of Implementation Rubric}
\label{sec:ease-of-implementation-rubric}

In this section, we describe how we ranked our test systems based on the expressiveness metric of \textit{Ease of Implementation}. Each author individually rated the test systems they developed the interfaces for, on the following criteria along a 5 point Likert scale:

\begin{enumerate}
    \item How easy was it to find the documentation for the test system and its driver on the system website?
    \item How comprehensive is the documentation? Does it provide examples?
    \item How easy was to find resources for help on forums like stack overflow or the system's own forum?
    \item How quick are the developers to respond to queries? How useful or helpful are their responses?
\end{enumerate}

This was followed by a round of discussion and reconciliation among the authors to arrive at the final ratings, which we discuss in the paper.



\end{document}